\documentclass[twocolumn,showpacs,floatfix,prl]{revtex4}
\usepackage{amsmath}
\usepackage{epsf}
\usepackage{graphicx}
\usepackage{dcolumn}
\usepackage{bm}

\begin{document}

\title{Nonanalytic corrections to the specific heat of a three-dimensional Fermi
liquid}
\author{Andrey V. Chubukov$^{1,2}$, Dmitrii L. Maslov$^{3}$, and Andrew J. Millis$^{4}$}
\date{\today}

\begin{abstract}
We revisit the issue of the leading nonanalytic corrections to the
temperature dependence of the specific heat coefficient, $\gamma (T)=C(T)/T,$
for a system of interacting fermions in three dimensions.
We show that the leading temperature
dependence of the specific heat coefficient $\gamma (T)-\gamma (0) \propto T^3 \ln T$
comes from two physically distinct processes.  The first process involves a thermal excitation of a
single particle-hole pair, whose components
 interact via a nonanalytic dynamic vertex. The second process
 involves an excitation of three particle-hole
pairs  which interact via the analytic static fixed-point vertex.
 We show that the single-pair
 contribution  is expressed via the backscattering
 amplitude of quasiparticles at the Fermi surface. The
 three-pair contribution  does not have
a simple expression in terms of scattering in particular directions.
We clarify the relation between these results and previous
 literature on both 3D and 2D systems, and
 discuss the relation between the nonanalyticities in $\gamma$
and those in spin susceptibilities.
\end{abstract}

\affiliation{ $^{1}$Department of Physics, University of Maryland,
College Park, MD 20742-4111\\
$^{2}$ Department of Physics, University of Wisconsin, 
1150 University ave., Madison, WI 53706.\\
$^{3}$
Department of
Physics, University of Florida, P. O. Box 118440, Gainesville, FL
32611-8440 \\
$^4$ Department of Physics, Columbia University, 538 W. 120th St.,New York, NY, 10027}
\pacs{71.10.Ay,71.10.Pm}
\maketitle
\section{Introduction}

The thermodynamic properties of itinerant fermionic systems is a subject of
long-standing experimental \cite{Abel66,Greywall83,Stewart94} and theoretical \cite
{LL,Eliashberg60,Anderson65,Balian65,Engelsberg66,
Doniach66,berk,brinkman,larkin,Amit68,Bealmonod,Riedel69,Pethick73a,Pethick75,Carneiro77,Coffey93,
mar'enko,belitz,Houghton98,fradkin,chitov_millis,chubukov03,galitski03,
Chubukov04,Chubukov05,galitski05,betouras05}
interest. It is generally accepted \cite{AGD,Pines66} that the low-temperature
behavior of a wide class of interacting fermionic systems is controlled by
the Fermi-liquid fixed point. This implies that the leading temperature
dependences are the same as for free fermions, but with renormalized
parameters. However, the first subleading corrections may differ
qualitatively from those of non-interacting fermions. This phenomenon was
first noticed in the context of the specific heat coefficient, $\gamma
(T)=C(T)/T$. For non-interacting fermions, $\gamma (T)$ has a regular
expansion in powers of $T^{2}$ about $T=0$, so the leading temperature
dependence is $\gamma (T)-\gamma (T=0)\sim T^{2}$. For interacting fermions,
$\gamma (T)$ is not an analytic function of $T^{2}$; the leading temperature
dependence is instead proportional to $T^{2}\ln (T)$ in three dimensions
(3D) and
$T$ in 2D
The $T^{2}\ln T$ term was first found by Eliashberg in a
theoretical study of
electrons interacting with acoustic phonons \cite{Eliashberg60}, and the
possibility of a nonanalytic temperature dependence of $\gamma$ was
subsequently (but apparently independently)
inferred from measurements of $\gamma $ for $^{3}$He by Abel, Wheately and Andersen \cite{Abel66}.
The $^3$He measurements led to a large literature~ \cite
{LL,Eliashberg60,Anderson65,Balian65,Engelsberg66,
Doniach66,berk,brinkman,larkin,Amit68,Bealmonod,Riedel69,Pethick73a,Pethick75,Carneiro77,Coffey93,
mar'enko,belitz,Houghton98,fradkin,chitov_millis,chubukov03,galitski03,
Chubukov04,Chubukov05,galitski05,betouras05}
which we review in detail below.
We note here a few crucial (and seemingly contradictory) results.
{\it (i)} In a very important paper \cite{Pethick73a}, Pethick and Carneiro showed
that for a 3D Fermi liquid the $T^2\ln T$ term in $\gamma$ was a Fermi-liquid effect, associated with
a combination of the multiple scattering of particle-hole pairs with small total
momentum and a particular behavior
of the quasiparticle interaction function $f_{{\bf p,p+q}}$ (also at small $q$).
{\it (ii)} More recently, it has been demonstrated~\cite{Chubukov04}  that in
 2D, the
 prefactor of the $T$ term in  $\gamma (T)-\gamma (T=0)\sim T$ is
determined entirely by  the squares of charge and spin components of
the \emph{backscattering} amplitude. [By ``backscattering'' we mean
scattering of fermions with almost opposite momenta. Transferred momentum
can be either small or near $2k_{F}.]$ On the other hand
the previous 3D work
\cite{Amit68,Pethick73a,Pethick75} found no special role of backscattering
for the $T^{2}\ln T$ nonanalyticity.
{\it (iii)}  The older work left the impression that
in 3D the nonanalyticities were particular to $\gamma$ and did not contribute to susceptibilities,
whereas more recent work (based mainly on perturbative calculations) has demonstrated
that nonanalytic corrections to the spin susceptibility occur both in
3D \cite{belitz} and in 2D \cite{chitov_millis,chubukov03,galitski03,Chubukov04,Chubukov05,galitski05,betouras05}.

In this paper we clarify the relation between
nonanalyticities in $\gamma $ for 3D and 2D systems and make a
few remarks concerning the relation between the nonanalyticities in $\gamma$
and those in susceptibilities.   We demonstrate that
backscattering plays a special role also in 3D, in the sense that
a part of the $T^{2}\ln
T $ term comes entirely from backscattering. This backscattering
contribution evolves smoothly between 2D and 3D, and is entirely responsible
for the nonanalytic part of $\gamma $ for $D<3$. In $D\ge3$, however, there
exists another, physically distinct, contribution to the $T^{2}\ln T$ term in
$\gamma $. This contribution does not occur for $D<3$
and is not expressible solely in
terms of the backscattering amplitude.
We argue that this contribution is important for the 3D spin susceptibility
as well.
We also clarify the relation between the expressions for the prefactor of
the $T^{2}\ln T$ term via the forward-scattering interaction in the
Pethick-Carneiro approach (and in subsequent analyses based on bosonization
\cite{Houghton98,fradkin}) and via the backscattering amplitude.
In particular, Pethick and Carnerio expressed the nonanalyticities
in terms of the small $q$ form of the scattering amplitude
$f_{\mathbf{p},\mathbf{p}+\mathbf{q}}$ for momenta slightly
displaced
from the Fermi surface. As we will
demonstrate explicitly, their result can be re-expressed
in terms of a dynamical interaction
between quasiparticles  at the Fermi surface, in which backscattering
plays a crucial role. At the same time, we show, in disagreement with  ~Refs.
\cite{Amit68,Coffey93} that processes involving
forward-scattering between quasiparticles at the Fermi surface do not
contribute to the nonanalyticity in $\gamma$
although forward scattering does make a nonanalytic contribution to
the self energy.

The remainder of this paper is organized as follows. In section II, we
outline the qualitative physics underlying the results presented here. In
section III, we present the formalism and introduce forward-scattering and
backscattering amplitudes. Sections IV and V give detailed calculations
of the nonanalyticity in  the entropy for the cases of
three and two spatial dimensions, respectively. Section VI presents
presents a brief discussion of the spin susceptibility. Section VII
explicates the relation of our results to previous calculations
of the self-energy, and section VIII compares our results for
the entropy to those
obtained in prior work. Section IX presents a brief comparison of our
results to the specific heat data for $^{3}$He.
Section X presents summary
and conclusions. An explicit calculation of the $2k_{F}$ contribution to the
nonanalyticity in the specific heat is presented in Appendix A.

\section{Qualitative Physics}

\begin{figure}[tbp]
\begin{center}
\epsfxsize=1.0\columnwidth
\epsffile{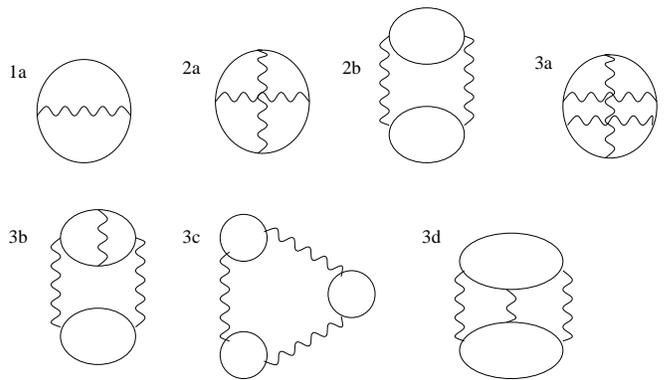}
\end{center}
\caption{
Skeleton 
diagrams for the thermodynamic potential.
The solid lines represent electron propagators, and the dashed lines
 represent the interaction potential.}
\label{Fig:Omegas}
\end{figure}

In this section we present a qualitative discussion of the physics
underpinning our key results. Subsequent sections provide the formal
justification. To compute the low-$T$ specific heat of a Fermi Liquid,
one expands the thermodynamic potential ($\Xi $) in the number of
particle-hole pairs excited above the Fermi-liquid ground state. Each
excited quasiparticle interacts with the corresponding excited quasihole and
with other particle-hole pairs via interaction vertices, $\Gamma $. Some low
order diagrams are shown schematically in Fig. \ref{Fig:Omegas}.

To obtain the entropy
$S=-d\Xi /dT$ we must average over all excited particle-hole
pairs, weighting the contributions by the derivative of the Bose
distribution function $\partial n_{B}\left( \varepsilon \right) /\partial T,$
which is odd in $\varepsilon$ and, at low temperature $T$
ensures that all excited electrons are near the Fermi surface.
The contributions also involve  the spectral function of
the quasiparticle-quasihole pairs [the imaginary part of the particle-hole
propagator $\Pi(\varepsilon,q)$], which is also odd in $\varepsilon$, and,
critically, is not an analytic function of $\varepsilon$.
For example, in  $D=3$,  
$\text{Im}\Pi (\varepsilon, q) \sim \left(\varepsilon/v_Fq\right)\Theta(v_Fq/\varepsilon-1)$
with  $\Theta (x)$ the  step function.
Therefore, only states with an odd number of
excited pairs contribute to the entropy, and thus to $\gamma (T)=dS/dT$. The
leading term has only one excited particle-hole pair (diagram 1a in Fig.\ref
{Fig:Omegas}) and for momentum-independent interaction $\Gamma$
yields for the interaction-dependent part of the entropy
\begin{equation}
\delta S(T)\propto \Gamma \int d\varepsilon ~\varepsilon~\frac{\partial n_{B}(\varepsilon )}{%
\partial T}\int \frac{d^{3}q}{q}.
\label{Sexample}
\end{equation}
The momentum integral in Eq [\ref{Sexample}]
is dominated by large $q\sim k_{F}$, and so yields a constant, whereas the
frequency integral yields $\delta S(T)\propto T$, i.e., $\gamma =$const. This is
the leading Fermi-liquid result.

The subleading terms contain contributions from three, five, etc thermally
excited particle-hole pairs. For thermally allowed excitations $\varepsilon
\sim T$ , so each extra particle-hole pair brings an extra factor of $T$ to
the thermodynamic potential. One might therefore expect an expansion of the
form $\delta S=\gamma _{0}T+\alpha T^{3}+_{\cdots }.$ However,
the non-analytic behavior of the particle-hole spectral function
implies that this conclusion is not correct. The non-analyticity enters in two ways.
 First, processes involving three pairs with the same
total momentum $q$ (see, e.g., Diagram 3c of Fig.\ref{Fig:Omegas})
contain the cube of the spectral function, so give a term
proportional to $(\varepsilon /q)^{3}$; the integral over $q$
leads to a logarithmic divergence cut off at $\varepsilon \sim T$
, and hence, in 3D,  to a $T^{3}\ln T$ contribution $\delta S(T)$.
Second, the general form of the interaction vertex $\Gamma $ has a term of order
$(\varepsilon /q)^{2}$ arising from the Kramers-Kronig transform
of Im$\Pi (\varepsilon ,q)$ which, in combination with the
$\varepsilon /q$ from a single excited particle-hole pair,
produces in $3D$ an additional $T^{3}\ln T$ contribution to
$\delta S(T)$ (see, e.g., diagram $2b$ in Fig. \ref{Fig:Omegas}).

Note that dimension $D=3$ is marginal in the sense that only for $D\leq 3$
the leading nonanalytic term is larger than the leading
analytic $\mathcal{O}%
(T^{3})$ correction. As will be seen, it is also marginal  in the sense that
for $D<3$ only the single-pair term contributes to the nonanalyticity.

Both contributions were identified by Pethick and Carneiro \cite{Pethick73a}.
They can be also clearly seen in the ``paramagnon model'' \cite
{Doniach66,larkin} of electrons coupled to overdamped spin
fluctuations (although only the three-pair contributions were discussed
in \cite{Doniach66,larkin}). The paramagnon part of the
entropy is obtained by summing up loop diagrams with a small momentum
transfer and is given by~\cite{Doniach66}
\begin{eqnarray}
\delta S &=&\frac{3}{8\pi T^{2}}\int d\varepsilon \frac{\varepsilon }{\sinh
^{2}\left( \varepsilon /2T\right) }  \notag \\
&&\times \int \frac{d^{D}q}{(2\pi )^{D}}~\tan^{-1} \frac{ 2g \text{Im}\Pi
(\varepsilon ,q)}{1+ 2g\text{Re}\Pi (\varepsilon ,q)},  \label{chu_1}
\end{eqnarray}
where $g$ is the coupling constant in the paramagnon model (see
Sec. \ref{sec:ZS} below).
One obtains one $\left( \varepsilon /q\right) ^{3}$ term from
the process involving three real excited particle hole pairs by
treating $\text{Re}\Pi$ as a constant and expanding the $\tan^{-1}$
to third order in $\text{Im}\Pi\sim \varepsilon/q$. In addition,
one obtains another $\left( \varepsilon /q\right) ^{3}$ term by
combining one power of Im$\Pi \propto \varepsilon /q$
(representing a real excited particle-hole pair) with the
nonanalytic term in Re$\Pi \propto \left( \varepsilon /q\right)
^{2}$ (representing the nontrivial dynamic structure arising from
$\text{Re}\Pi$).

The two processes evidently  represent different physics and differ mathematically
as well. The
single-pair mechanism occurs already at the second-order in the interaction $g$
and is not specific to $D=3$. A simple extension of the $3D$ analysis to
arbitrary integer $D$ shows that the single-pair mechanism
yields a non-analytic $%
\delta S\propto T^{D}$ in any dimension (with an additional $\ln T$ in odd
dimensions). On the contrary, the three-pair mechanism occurs to third
order in the interaction and gives rise to non-analyticity in $\delta S$
only in odd dimensions $D\geq 3$.

Let us consider the kinematics of these contributions in more
detail. In both processes, the nonanalyticity arises from Landau
damping, which for the physically relevant limit of
$\varepsilon\ll q$ is dominated by fermions with momenta nearly
perpendicular to ${\bf q}$. For a
single-pair process, represented, e.g., by diagram
(Fig.~\ref{Fig:Omegas} 2b), the constraint that the momenta on all
four fermion lines  are perpendicular to the same vector ${\bf q}$
implies that in two dimensions all four momenta are either nearly parallel to each
other or two are nearly antiparallel to the other two.  A closer
analysis \cite{Chubukov04} shows that only the nearly antiparallel
case contributes to the nonanalyticity in $\gamma$ so the
nonanalyticity is controlled by backscattering. However,  it
appears surprising that backscattering would play a special role
in  $3D$. Indeed, in $D=3$ the constraint that all four momenta
are perpendicular to $q$ implies only that all four momenta lie in
a common plane, so that any multiparticle excitation would seem to
involve a more general scattering process involving four fermions,
two of which have momenta near $k$ and two near $p$ but with ${\bf
k}$ and ${\bf p}$ perpendicular to ${\bf q}$ but not otherwise
constrained. Indeed the three pair contributions have precisely
this structure. Nevertheless we will show explicitly that in the
single pair process only the case of nearly antiparallel momenta
contributes.

Some physical understanding of the importance of backscattering even in $3D$
may be obtained by considering the problem in a slightly more general way.
The discussion just given has focussed on ``small momenta''
in the sense that it considered processes describable in terms of electron-hole
pairs where the electron and the hole momenta are very close (on the scale set
by $k_F$), and the nonanalyticity arises from the singular structure of the
Landau damping $\sim \varepsilon/q$ in the small $\varepsilon,q$ limit. However,
 Fermi liquids also exhibit nonanalyticities involving momenta near $2k_F$,
(reflected for example in the slow decay of Friedel oscillations), and previous literature
has raised the question of the contribution of $2k_F$ processes to the nonanalytic
behavior~\cite{Amit68}.  To understand this issue it is useful to consider
 again the simple perturbative
contribution shown in diagram $2b$ of Fig. \ref{Fig:Omegas}, which may be written
as the product of two particle-hole bubbles:
\begin{equation}
\Omega_{2b}\sim \int d\varepsilon \frac{dn_B (\varepsilon)}{dT}
\int d^Dq\text{Im}
\Pi(q,\varepsilon)\text{Re}\Pi(q,\varepsilon). \label{omega2b}
\end{equation}
There are two possible momentum regions which may give rise to nonanalyticities:
small momentum and $q \sim 2k_F$.
Consider the contribution to $\Omega_{2b}$ arising from large momentum transfer,
${\bf q}=(2k_F+{\bar q}){\hat q}$. The nonanalyticity arises
from the ``Landau-damping'' structure of $\Pi$,
which in $3D$ and for momenta near $2k_F$
is  $\text{Im}\Pi({\bf q},\varepsilon) \propto \varepsilon~
 \Theta(|\varepsilon|-{\bar q})$
and $\text{Re}\Pi-\text{Re}\Pi(2k_F,0)\sim \varepsilon^2/{\bar q}$ (see Appendix A).
We note that the orientation of ${\bf q}$
is irrelevant, so $\int d^D q $ becomes a one-dimensional integral over the
scalar quantity ${\bar q}$ leaving a logarithm for $D=3$.. The $2k_F$ ``Landau damping'' terms involve
fermions with momenta close to $k_F$, so that if $q \approx 2k_F$ then in each bubble
one has one fermion of momentum ${\bf k} \approx {\bf q}/2$ and one with ${\bf k} \approx -{\bf q}/2$;
thus the $2k_F$ contribution involves a one-dimensional process controlled by
the ``backscattering amplitude'' $\Gamma({\bf k},-{\bf k};{\bf k},-{\bf k})$.
However, we may also view the diagram for $\Omega_{2b}$ in a different way,
regarding the two fermion lines with momentum ${\bf k}\sim{\bf q}/2$ as a bubble
with a small total momentum ${\bar q}$, and similarly with the two lines of momentum
near $-{\bf q}$. Thus for the process with one real excitation and
 first dynamical correction
to interaction, the $2k_F$ contribution may be subsumed into the small $q$ contribution.

The total nonanalytic contribution to $\delta S(T)$ may
thus be described in terms of
two sorts of small momentum processes. One may be thought
of in terms of $2k_F$ processes, and involves the vertex
$\Gamma (\mathbf{k},-\mathbf{k};-\mathbf{k},\mathbf{k})$.
The other arises from consideration of small $q$ processes
and seemingly involves   the effective
interaction $\Gamma (\mathbf{k},\mathbf{p};\mathbf{k},\mathbf{p})$
with arbitrary values of  ${\bf k}\cdot {\bf p}$.
Graphically, the two contributions differ only by interchanging the two
outgoing momenta in the effective interaction. It is thus plausible to
expect that both contributions are expressed in terms of the spin and charge
components of the same antisymmetrized vertex. The
vertex $\Gamma (\mathbf{k},-%
\mathbf{k};-\mathbf{k},\mathbf{k})$ is the spin part of the backscattering
amplitude. We may expect that $\Gamma (\mathbf{k},\mathbf{p};\mathbf{k},%
\mathbf{p})$ is expressed in terms of charge component of the same
amplitude. If this is the case, then $\mathbf{p}$ must be antiparallel to $%
\mathbf{k}$, i.e., the original small $q$, single-pair contribution also
involves 1D process, and hence the total single-pair contribution involves
1D scattering.

For the three-pair mechanism, this argument does not hold.
The kinematics of $2k_F$ processes is such that if more than
two pairs are involved the result does not have a strong enough nonanalyticity.
Indeed, combining three factors of Im$\Pi (\varepsilon ,2k_{F})\propto
\varepsilon $, we obtain only a regular term in $\delta S(T)$. Thus
the argument for the presence, in the calculation, of a purely one-dimensional
scattering process fails and backscattering plays no special role.

A principal aim of the present paper is to present a more rigorous
analysis substantiating the qualitative arguments given above. In this analysis
the effects of single-pair and three-pair processes are
separated and the contribution of each
to the nonanalyticities in $\gamma$ is determined
in two and three dimensions.

\section{Formalism}

We calculate the specific heat coefficient $\gamma (T)=C/T$ by evaluating
the difference between the thermodynamic potential, $\Xi ,$ and its value at
 $T=0$. The Luttinger-Ward expression for $\Xi $ is \cite{luttinger}
\begin{equation}
\Xi =-\mathrm{Tr}\left( \ln \left[ G_{0}^{-1}-\Sigma \right] +\Sigma
G\right) +\Xi _{\mathrm{skel}}  \label{Omegadef}
\end{equation}
with $\Sigma $ the exact self-energy (regarded as a functional of the full
Green function $G$) and $\Xi _{\mathrm{skel}}$ is the usual skeleton diagram
expansion for all interaction corrections to $\Xi $ that are not accounted
for by the first two terms. The trace is taken over space, time and spin
variables.

It is not necessary to obtain the nonanalytic contributions the self-energy $%
\Sigma $ in order to evaluate the nonanalytic contributions to $\Xi $. For
what appear to be historical reasons associated
 with the fact that
nonanalytic contributions to $\Sigma $ were a focus of many  earlier studies,
the previous literature evaluates the Tr$ \ln \left[ G_{0}^{-1}-\Sigma %
\right] $ separately from the other contributions, thereby
effectively adding and subtracting contributions arising from
nonanalyticities in $\Sigma $. However,
the stationary of $\Xi$ with respect to variations of $\Sigma$ implies that
 it is in fact sufficient to consider only the
skeleton diagram $\Xi _{\text{skel}}$, which may be evaluated using the
leading renormalized quasiparticle Green function $G_{\text{qp}}=Z^{-1}\left[
\omega -v_{F}^{\ast }(k-k_{F})\right] ^{-1}$, where $Z$ is the
renormalization factor and $v_{F}^{\ast }$ is the renormalized Fermi
velocity. We emphasize  that, in physical terms, this corresponds to
expanding $\Xi $ in the number of particle-hole pairs thermally excited
above the ground state, but assuming that these interact according to the $T=0$
Landau Fermi Liquid fixed-point Hamiltonian.

A further subtlety arises: even if the Green function takes the
fixed-point form, the $2k_F$ particle-hole bubble will have a
nonanalytic temperature dependence.  However, as we discussed in
the previous section, singular $2k_{F}$ scattering can be
re-expressed as small momentum scattering, and is fully accounted
for if we restrict to small $q$ but consider wavy lines in Fig.
\ref{Fig:Omegas} as antisymmetrized interactions, i.e., as
fixed-point vertices. The small-$q$ diagrams have nonanalyticities
arising from the Landau damping $\varepsilon/v_Fq$ which has
negligible temperature dependence. Once the restriction to small
$q$ is accomplished, we may calculate the leading low-$T$ behavior
of $\Xi (T)-\Xi (0)$ by focusing
only on the $T$-dependence arising from the difference between $%
T\sum_{\omega _{n}}$ and $\int d\omega /2\pi $, and
 neglecting any explicit
temperature dependence of $G$, $\Sigma $ or interaction vertices. These
latter contributions give only analytic, $\mathcal{O}\left( T^{2}\right) $
corrections to $\gamma $. Appendix A confirms this argument by presenting an
explicit calculation of the contribution to $S$ arising from $2k_F$ processes.

One can employ two strategies to calculate nonanalytic terms in $\Xi \left(
T\right) .$ The first one is to evaluate the thermodynamic potential
directly in Matsubara frequencies $\varepsilon _{m}$. This strategy was
adopted in recent study of 2D systems\cite{Chubukov04}. A
drawback of this approach is that it does not distinguish between the
real and imaginary parts of $\Pi (\varepsilon,q)$, and
the physical picture of excited particle-hole pairs does not arise. The
second --and more intuitive--approach, adopted in this paper, is to work in
real frequencies, when Im$\Pi \propto \varepsilon $ and Re$\Pi \propto
\varepsilon ^{2}$ have different physical meaning: the former describes real
particle-hole pairs with given momentum and energy, whereas the latter
describes the interaction arising from virtual pairs.

To evaluate non-analytic terms in $\Xi $ for a generic Fermi Liquid we need
two quantities. The first one is the propagator $\mathcal{P}_{\text{ph}%
}(\varepsilon ,q;\mathbf{n}_{k}),$ describing a particle-hole pair with
small total momentum $q$ and small energy $\varepsilon ,$ respectively, and
with the particle and hole momenta near the Fermi surface and in direction $%
\mathbf{n}_{k}.$ The second one is the fully renormalized, particle-hole
\textit{reducible} vertex $\Gamma _{\alpha \beta ;\gamma \delta
}(\varepsilon ,q;\mathbf{n}_{k},\mathbf{n}_{p})$ describing scattering of
one particle-hole pair state into another.

For small $q$ and $\varepsilon $, $\mathcal{P}_{\text{ph}}$ may be written
as a sum of two terms: $\mathcal{P}_{\text{ph}}=\mathcal{P}_{\mathrm{0}}+%
\mathcal{P}$. The analytic part $\mathcal{P}_{\mathrm{0}}$ is a function of $%
\varepsilon $ and $q^{2}$ and contains contributions from virtual processes
involving states both near and far from the Fermi level. The nonanalytic
part, denoted by $\mathcal{P}$, is determined solely by the properties of
Fermi surface states and depends only on $z=v_{F}^{\ast }q/\varepsilon $.
Expanding the product of two quasiparticle propagators near the Fermi
surface, we obtain
\begin{equation}
\mathcal{P}(z;\mathbf{n}_{k})=\frac{1}{S_{D}}\frac{z^{-1}}{z^{-1}-\mathbf{n}%
_{k}\cdot \mathbf{n}_{q}}  \label{PiNA}
\end{equation}
with $S_{D}$ being the surface area of the sphere of unit radius in $D$
dimensions. We absorb the factor of inverse velocity and the
factors of quasiparticle weight into the definition of $\Gamma ,$ and we
have specialized to rotationally invariant systems. The frequency-dependent
part of the polarization bubble $\Pi (\varepsilon ,q)$ is obtained by
averaging $\mathcal{P}(\varepsilon ,q;{\bf n}_{k})$ over $\mathbf{n}_{k}$:
\begin{equation*}
\Pi (z)=\int d\mathbf{n}_{k}\mathcal{P}(z;\mathbf{n}_{k}).
\end{equation*}
In 3D,
\begin{eqnarray}
\text{Re}\Pi (z) &=&\frac{1}{2 z}\ln \left| \frac{z+1}{z-1}\right| ,
\label{re} \\
\text{Im}\Pi \left( z\right)  &=&-\frac{\pi }{2 z}\Theta \left( \left|
z\right| -1\right) .
\end{eqnarray}
In the perturbation theory, when the interaction depends only on the
momentum transfer but not on the incoming momenta, one needs to know only
the angle-averaged bubble. However, the interaction in a generic Fermi
Liquid is described by a vertex $\Gamma _{\alpha \beta ;\gamma \delta
}(\varepsilon ,q;\mathbf{n}_{k},\mathbf{n}_{p}),$ which depends not only on
the momentum transfer $q$ but also on the relative
 directions of the incoming momenta
$ \mathbf{n}_{k}\text{ and }\mathbf{n}_{p}.$  Therefore we need
 need the $\mathbf{n}_{k}$-dependent propagator
 $\mathcal{P}(\varepsilon ,q;\mathbf{n}_{k})$. In evaluating $\Xi ,$ we will then have to perform
angular integrations of the convolutions of $\mathcal{P}$ and $\Gamma $.%

The fixed-point vertex $\Gamma _{\alpha \beta ;\gamma \delta }(\varepsilon
,q;\mathbf{n}_{k},\mathbf{n}_{p})$ is an (anti)symmetrized sum of all
scattering processes between fermions \textit{at the Fermi surface}. We
may neglect any analytic dependence of $\Gamma $ on $\varepsilon $ and $q$
separately, and consider $\Gamma $ to be a function only of $z$.
 The fixed point vertex is also
 a tensor in the spin space. It is convenient to express this tensor
 as a sum of charge and spin components
\begin{eqnarray}
\Gamma _{\alpha \beta ;\gamma \delta }(z;\mathbf{n}_{k},\mathbf{n}_{p})
&=&\delta _{\alpha \gamma }\delta _{\beta \delta }\Gamma _{c}(z;\mathbf{n}%
_{k},\mathbf{n}_{p})  \notag \\
&&+\mathbf{\sigma _{\alpha \gamma }\cdot \sigma }_{\beta \delta }\Gamma
_{s}(z;\mathbf{n}_{k},\mathbf{n}_{p}),  \label{spincharge}
\end{eqnarray}
where $c$ and $s$ refer to the charge and spin sectors, respectively.
In a model with bare fermion-fermion
interaction $U(q)\rho(q)\rho(-q)$ to first order in the interaction we have
\begin{equation}
\Gamma _{\alpha \beta ;\gamma \delta }(z;\mathbf{n}_{k},\mathbf{n}_{p})
=\delta_{\alpha\gamma}\delta_{\beta\delta}u(0)
-\delta_{\alpha\delta}\delta_{\beta\gamma}u(k_{F}|\mathbf{n}%
_{k}-\mathbf{n}_{p}|)
\end{equation}
and thus
\begin{eqnarray}
\Gamma _{c}(z;\mathbf{n}_{k},\mathbf{n}_{p}) &=&u(0)-(1/2)u(k_{F}|\mathbf{n}%
_{k}-\mathbf{n}_{p}|)
\label{gammacfirstorder}\\
~~\Gamma _{s}(z;\mathbf{n}_{k},\mathbf{n}_{p}) &=&-(1/2)u(k_{F}|\mathbf{n}%
_{k}-\mathbf{n}_{p}|),
\label{gammasfirstorder}
\end{eqnarray}
where $u(q)\equiv U(q)k_{F}^{2}/(\pi ^{2}v_{F}^{\ast }).$

Pethick and Carneiro \cite{Pethick73a} showed that the charge and spin
components of $\Gamma $ contribute independently to
the nonanalyticity in $\Xi $%
, i.e., $\Xi _{\mathrm{NA}}=\sum_{a}w_{a}\Xi _{a}$, where $a=c,s$ and $%
w_{c}=1,~w_{s}=3$ (for $SU\left( 2\right) $ fermions). We explicitly
verified and confirmed their result, which indeed follows simply from
an  elementary consideration of
the possible distributions of charge and spin vertices along
fermionic loops. To make our presentation more compact,
we will consider only the charge component of $\Gamma $ and omit the index $c$.
The spin component will be restored in the final results.

The forward-scattering and backscattering processes in these notations
correspond to $\Gamma (z;\mathbf{n}_{p},\mathbf{n}_{p})$ and
$\Gamma (z,\mathbf{n}_{p},-\mathbf{n}_{p})$, respectively. In the first process, all
four fermionic momenta are almost equal; in the second one,
the two incoming and the two outgoing
 momenta are almost antiparallel.

The vertex $\Gamma (z;\mathbf{n}_{p},\mathbf{n}_{k})$ satisfies the integral
equation
\begin{eqnarray}
&&\Gamma (z;\mathbf{n}_{k},\mathbf{n}_{p})=\Gamma ^{k}(\mathbf{n}_{k}\cdot
\mathbf{n}_{p})  \notag \\
&&+\int d\mathbf{n}_{l}\Gamma ^{k}(\mathbf{n}_{k}\cdot \mathbf{n}_{l})%
\mathcal{P}(z;\mathbf{n}_{l})\Gamma (z;\mathbf{n}_{l},\mathbf{n}_{p}),
\label{vertex}
\end{eqnarray}
where the integral is over the area of the unit sphere in $D$ dimensions.
Virtual processes which contribute to $\mathcal{P}_0$ are absorbed
into the ``bare'' vertex $\Gamma ^{k}(\mathbf{n}_{k}\cdot \mathbf{n}%
_{p})=\Gamma (\infty ;\mathbf{n}_{l},\mathbf{n}_{p})$. The notation $\Gamma
^{k}$, borrowed from ~Ref.\cite{LL}, comes from the observation that $%
\mathcal{P}$ vanishes at $z=\infty $, i.e., at $\varepsilon /v_{F}^{\ast
}q\rightarrow 0$. We recall that $\Gamma ^{k}$ coincides, up to the $Z-$%
factor, with the quasiparticle scattering amplitude. The scattering
amplitude differs from the Landau function, which is related to
 $\Gamma ^{\omega }\left( \mathbf{n}%
_{k}\cdot \mathbf{n}_{p}\right) =\Gamma (z=0,\mathbf{n}_{k},\mathbf{n}_{p})$.
The scattering amplitude may be expanded in spherical (or, in two
dimensions, circular) harmonics as
\begin{equation}
\Gamma ^{k}(\mathbf{n}_{k},\mathbf{n}_{p})=\Gamma ^{k}(\mathbf{n}_{k}\cdot
\mathbf{n}_{p})=\sum_{L}{\bar{\Gamma}}_{L}P_{L}(\mathbf{n}_{k}\cdot \mathbf{n%
}_{p}).  \label{leg}
\end{equation}
Here $P_{L}$ are Legendre polynomials in $D=3$ and cosines in $D=2$. The
partial amplitudes, ${\bar{\Gamma}}_{L},$ of $\Gamma ^{k}$ are related to
the partial amplitudes, $\Gamma _{L},$ of $\Gamma ^{\omega }$ via
\begin{equation*}
{\bar{\Gamma}}_{L}=\frac{\Gamma _{L}}{1+\Gamma _{L}(2L+1)^{-1}}
\end{equation*}
and
\begin{equation*}
{\bar{\Gamma}}_{L}=\frac{\Gamma _{L}}{1+\Gamma _{0}(2-\delta _{L,0})^{-1}}
\end{equation*}
in 3D and 2D, respectively.

\begin{figure}[tbp]
\begin{center}
\epsfxsize=0.7\columnwidth
\epsffile{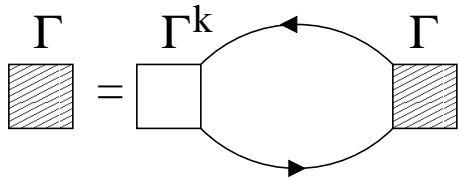}
\end{center}
\caption{Graphical representation of Eq. (\ref{vertex}). }
\label{Fig:Gamma}
\end{figure}

To generate an expansion in a number of particle-hole pairs we need an
auxiliary vertex, $\bar{\Gamma}(z;\mathbf{n}_{k},\mathbf{n}_{p})$, which
satisfies the same equation as in Eq.(\ref{vertex}) but with Re$\mathcal{P}%
(z;\mathbf{n}_{l})$ instead of full $\mathcal{P}(z;\mathbf{n}_{l})$%
\begin{eqnarray}
&&\bar{\Gamma}(z;\mathbf{n}_{k},\mathbf{n}_{p})=\Gamma ^{k}(\mathbf{n}%
_{k}\cdot \mathbf{n}_{p})  \notag \\
&&+\int d\mathbf{n}_{l}\Gamma ^{k}(\mathbf{n}_{k}\cdot \mathbf{n}_{l})\text{%
Re}\mathcal{P}(z;\mathbf{n}_{l})\bar{\Gamma}(z;\mathbf{n}_{l},\mathbf{n}%
_{p}).
\end{eqnarray}

\begin{figure}[tbp]
\begin{center}
\epsfxsize=0.5\columnwidth
\epsffile{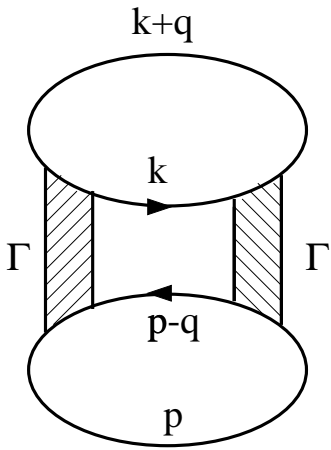}
\end{center}
\caption{Diagram {\text 2b} for the thermodynamic potential with fully dressed
vertices $\Gamma ({\bf q}, {|bf k} - {\bf p})$,
 instead of the bare interaction $U(q)$}
\label{Fig:2bgamma}
\end{figure}
The leading low-$T$ contributions to the entropy involve particle-hole pairs
excited above the Fermi surface and can be mathematically described by
combinations of Im\textbf{$\mathcal{P}$}, representing the excited
particle-hole pairs, and vertices $\bar{\Gamma}$, describing the
interactions between these pairs. The diagrammatic expansion of the
 thermodynamic potential in series of particle-hole pairs is 
 obtained from the skeleton diagrams of Fig. \ref{Fig:Omegas} by 
 replacing the
wavy lines by fully dressed interaction vertices. For example, diagram
\textit{2b} of Fig. \ref{Fig:Omegas} is replaced by the diagram in 
Fig.~\ref{Fig:2bgamma}. An important new feature of this expansion
 is that the vertices now depend
 not only on the momentum transfer (${\bf q}$) but also on the
incoming momenta (${\bf k}$ and ${\bf p}$). 
 Consideration of diagrams involving excitations and multiple
scattering of pairs with total energy $\varepsilon $ and momentum $q$ leads
to an expression for the nonanalytic contribution to the entropy per unit
volume $\delta S_{\text{NA}}=-(1/V)d\Xi _{\text{skel}}/dT$ of the form
\begin{equation}
\delta S_{\text{NA}}=\int \frac{d\varepsilon }{\pi }\int \frac{d^{D}q}{(2\pi )^{D}}%
\frac{\varepsilon }{4T^{2}\sinh ^{2}\left( \frac{\varepsilon }{2T}\right) }%
\Phi (\varepsilon ,q),  \label{S}
\end{equation}
where
\begin{eqnarray}
&&\Phi (\varepsilon ,q) =\Phi (z)=\sum_{l}\frac{(-1)^{l+1}}{2l} \times 
\notag \\
  &&\left[ \prod_{j=1}^l \int {d \mathbf{n}_{p_{j}}} \text{Im}\mathcal{P}(z,\mathbf{n}_{p_{j}})
\Lambda (z,\{\mathbf{n}_{p_{j}}\})\right].  \label{phi}
\end{eqnarray}
Here the factors of Im$\mathcal{P}$ represent the excited particle-hole
pairs and the factors of $\Lambda $ (determined by combinations of the
$\bar{\Gamma}$ from Eq. \ref{vertex}) describe the interaction between them. We
note that because Im$\mathcal{P}$ is odd in $\varepsilon $ whereas  $\bar{%
\Gamma}$ is even, only terms involving odd powers of Im$\mathcal{P}$, i.e.,
odd number of pairs, contribute to $\delta S_{\text{NA}}$. To extract the
nonanalytic behavior it is convenient to introduce $x=\varepsilon /2T$ and
rewrite Eq. (\ref{S}) in dimension $D$ as
\begin{eqnarray}
\delta S_{\text{NA}} &=&\frac{2^{D+1}S_{D}T^{D}}{(2\pi )^{D}(v_{F}^{\ast })^{D}}%
\int_{0}^{\infty }\frac{dx}{\pi }\frac{x^{D+1}}{\sinh ^{2}x}  \label{SNA} \\
&&\times \int_{1}^{v_{F}^{\ast }k_{F}^{{}}/Tx}z^{D-1}dz\Phi (z). \nonumber
\end{eqnarray}
Here, we used the fact that Im$\mathcal{P}(z,\mathbf{n}_{p})=0$ if $\left|
z\right| <1$.

\section{Non-analyticity in Specific Heat in Dimension D=3}
\label{sec:D3}

\subsection{Overview}

This section presents results for the nonanalytic entropy contribution $\delta S_{%
\text{NA}}\propto T^{3}\ln T$ for a three- dimensional Fermi
Liquid. After an overview we present results based on  expanding
Eqs.(\ref{vertex}, \ref{SNA}) in the number of physical excited
particle-hole pairs. We show that $\delta S_{\text{NA}}$ can be
expressed in terms of series of partial amplitudes $\Gamma _{L}$;
although a compact, useful closed-form expression exists only for
a part of $\delta S_{\text{NA}}$ corresponding to a single-pair
mechanism. In the last subsection we consider a toy model in which
the expansion in particle-hole pairs can be carried out to all
orders.

The key point of the calculation is this: the requirement of odd powers
means that in Eq.(\ref{phi}) only terms with odd $j$, i.e., odd powers of Im$\mathcal{P}(z,%
\mathbf{n}_{p_{j}})$ contribute to $\Phi (z)$ . At large $z$%
, Im$\mathcal{P}$ is of order $z^{-1}$; while $\Lambda =A+Bz^{-2}+O(z^{-4})$
is finite at $z^{-1}=0$. The leading large-$z$ contribution is thus $%
\mathcal{O}\left( z^{-1}\right) $; the integral in Eq. (\ref{SNA}) is
dominated by the upper limit and gives a contribution to the $z$-integral $%
\sim T^{-2}+$const; the net result is contributions of order $T$ and $T^{3}$
to the entropy which are not of interest here. The subleading contribution
come from the $j=3$ term in Eq.(\ref{phi}) and from the $z^{-2}$ term in $%
\Lambda $ combined with the $j=1$ term in Eq [\ref{phi}].
Both give contributions of order $z^{-3}$ to $\Phi (z)$.

Expanding $\Phi(z)$ asymptotically for large argument yields $\Phi
\left( z\right) $ at large $z$ as
\begin{equation}
\Phi (z\rightarrow \infty )=\frac{\Phi _{1}}{z^{{}}}+\frac{\Phi _{3}}{z^{3}}%
+...,  \label{phiexpand}
\end{equation}

Subtracting off the leading term, inserting the result in Eq. (\ref{SNA}%
), and integrating gives for the entropy per unit volume
\begin{equation}
\delta S_{NA}=\frac{4\pi ^{3}}{5}\Phi _{3}\left( \frac{T}{v_{F}^{\ast }}\right)
^{3}\ln \left( \frac{v_{F}^{\ast }k_{F}}{T}\right) .  \label{S3dfinal}
\end{equation}
The problem is therefore to calculate $\Phi_3$.
It is convenient to split $\Phi _{3}$ into contributions from processes with
one and with three excited real particle-hole pairs, i.e.,
\begin{equation*}
\Phi _{3}=\Phi _{3}^{(1)}+\Phi _{3}^{(3)};
\end{equation*}
and to discuss the contributions separately.

\subsection{Expansion in the number of excited particle-hole pairs}

\subsubsection{Single particle-hole pair}

For the term involving only one excited particle-hole pair, reference to
Fig. \ref{Fig:Omegas}\textit{a} shows that the interaction term is the
quasiparticle vertex $\bar{\Gamma}$, from which we need the contribution
proportional to $z^{-2}$ . Hence
\begin{equation}
\Phi _{3}^{(1)}=\lim_{z\rightarrow \infty }\frac{z^{3}}{2}\int d\mathbf{n}%
_{p}\text{Im}\mathcal{P}(z;\mathbf{n}_{p})\Gamma (z;\mathbf{n}_{p},\mathbf{n}%
_{p})  \label{phi31}
\end{equation}
with $\Gamma (z;\mathbf{n}_{p})$ being the order $z^{-2}$ term from the
solution of Eq. (\ref{vertex}). Direct inspection shows that there are two
contributions to $\Gamma (z;\mathbf{n}_{p})$; one, denoted $\Gamma ^{(1)}$,
of first order in Re$\mathcal{P}$ and another one, denoted $\Gamma ^{(2)}$,
of second order. These contribution are give by
\begin{eqnarray}
&&\Gamma ^{(1)}(z)=\int d\mathbf{n}_{k}\left[ \Gamma ^{k}(\mathbf{n}%
_{p}\cdot \mathbf{n}_{k})\right] ^{2}\text{Re}\mathcal{P}(z,\mathbf{n}_{k}),
\label{gam1} \\
&&\Gamma ^{(2)}(z)=\int d\mathbf{n}_{k_{1}}d\mathbf{n}_{k_{2}}\Gamma ^{k}(%
\mathbf{n}_{p}\cdot \mathbf{n}_{k_{1}})\text{Re}\mathcal{P}(z,\mathbf{n}%
_{k_{1}})  \notag \\
&&\times \Gamma ^{k}(\mathbf{n}_{k_{1}}\cdot \mathbf{n}_{k_{2}})\text{Re}%
\mathcal{P}(z,\mathbf{n}_{k_{2}})\Gamma ^{k}(\mathbf{n}_{k_{2}}\cdot \mathbf{%
n}_{p}).  \label{gam2}
\end{eqnarray}
Accordingly, $\Phi _{3}^{(1)}$ is a sum of two contributions, $\Phi
_{3}^{(1,1)}$ and $\Phi _{3}^{(1,2)}.$ If $\Gamma ^{k}$ is
angle-independent, the angular integrations in Eqs.(\ref{gam1},\ref{gam2})
yield  $\Gamma^{(1)} (z) \sim (\Gamma ^{k})^{2}$Re$\Pi (z)\propto z^{-2}$ and
$\Gamma^{(2)} (z) \sim (\Gamma
^{k})^{3}\left[ \text{Re}\Pi (z)\right] ^{2}\propto z^{-4},$ respectively
[cf. Eq.(\ref{re})]. Hence $\Phi _{3}^{(1,1)} = O(1)$ and $\Phi
_{3}^{(1,2)}\propto z^{-2}\ll $ $\Phi _{3}^{(1,1)}.$
However, if $\Gamma ^{k}
$ is angle-dependent, $\Phi _{3}^{(1,1)}$ and $\Phi _{3}^{(1,2)}$are of the
same order. Indeed, the large-$z$ limit of Re$\mathcal{P}(z,\mathbf{n}_{k})$
in Eq.(\ref{PiNA}) is $\sim z^{-1}/\left( {}\mathbf{n}_{k}\cdot {}\mathbf{n}%
_{q}\right) $, which means that the product of two Re$\mathcal{P}(z,\mathbf{n%
}_{k})$ in Eq.(\ref{gam2}) is odd in $\mathbf{n}_{k_{1,2}}.$ However, this
oddness is compensated by a combination of angular harmonics of $\Gamma ^{k}$
 in Eq.(\ref{gam2}), which contains a factor of $({}\mathbf{n}_{k_{1}}\cdot {}%
\mathbf{n}_{q})({}\mathbf{n}_{k_{2}}\cdot {}\mathbf{n}_{q}).$ As a result,
the angular integrals are finite and $\Gamma ^{(2)}(z)\propto \left( \text{Re%
}\mathcal{P}\right) ^{2}\propto z^{-2},$ which is of the same order as $%
\Gamma ^{(1)}(z).$ One can readily verify that terms with three, four,
etc factors of Re$\mathcal{P}$ are irrelevant, as they scale at least as $%
z^{-3}$.

We consider the two contributions, $\Phi _{3}^{(1,1)}$ and $\Phi
_{3}^{(1,2)},$ separately, beginning with the first one.
A conventional way to proceed would be to expand $\Gamma (\mathbf{n}%
_{k}\cdot \mathbf{n}_{p})$ in  Legendre polynomials, as in (\ref
{leg}), and evaluate the integrals order by order in $\Gamma _{L}$. However,
it is more convenient to expand $\Gamma (\mathbf{n}_{k}\cdot \mathbf{n}_{p})$
in powers of $\mathbf{n}_{k}\cdot \mathbf{n}_{p}$, and to write out the
scalar product $\mathbf{n}_{k}\cdot \mathbf{n}_{p}$ explicitly in terms of
polar angles referring to the direction of $q$, i.e.,
\begin{eqnarray}
&&\Gamma ^{k}(\mathbf{n}_{k}\cdot \mathbf{n}_{p})=  \notag \\
&&\sum_{n}{\tilde{\Gamma}}_{n}\left( \cos \theta _{k}\cos \theta _{p}+\cos
\phi _{k}\sin \theta _{k}\sin \theta _{p}\right) ^{n}.  \label{gamma3d}
\end{eqnarray}
In these notations, the backscattering amplitude is
\begin{equation}
\Gamma _{\text{BS}}=\Gamma ^{k}(\mathbf{n}_{p}, (-\mathbf{n}%
_{p}))=\sum_{n}(-1)^{n}{\tilde{\Gamma}}_{n}.  \label{backscat}
\end{equation}
Inserting Eq. (\ref{gamma3d}) into Eq. (\ref{gam1}) and then the result into
Eq. (\ref{phi31}), recalling that Im$\mathcal{P}=-z^{-1}\delta (z^{-1}-%
\mathbf{n}_{p}\cdot \mathbf{n}_{q})/4\pi $, exploiting the azimuthal
symmetry and the reflection symmetry about the plane $\mathbf{n}_{p}\cdot
\mathbf{n}_{q}=0,$ and defining $y=\cos \theta _{k},$ yields
\begin{equation}
\Phi _{3}^{(1,1)}=-\frac{\pi }{4}~\sum_{nm}{\tilde{\Gamma}}_{n}{\tilde{\Gamma%
}}_{m}I_{nm},  \label{phi13}
\end{equation}
where
\begin{equation}
I_{nm}=\lim_{z\rightarrow \infty }z^{2}\int_{-1}^{1}dyJ_{nm}(z,y){\mathcal{P}%
}\frac{z^{-1}}{z^{-1}-y}  \label{imn}
\end{equation}
and
\begin{eqnarray}
J_{nm}(z,y) &=&\int \frac{d\phi }{2\pi }\left( \frac{y}{z}+\cos \phi \sqrt{%
1-z^{-2}}\sqrt{1-y^{2}}\right) ^{n+m}  \notag \\
&=&\sum_{L=0...n+m}\frac{(n+m)!}{(n+m-L)!L!}(\cos \phi )^{n+m-L}  \notag \\
&&\times \left( \frac{y}{z}\right) ^{L}\left( \sqrt{1-z^{-2}}\sqrt{1-y^{2}}%
\right) ^{n+m-L}.  \label{Jmn}
\end{eqnarray}
If $n+m=2P$ is even, then only terms with even $L$ contribute to the integral in Eq.(%
\ref{Jmn}), and $J$ is an even function of $y$. In this case, the principal
value integral gives a result of order $z^{-2},$ and only the $L=0$ term in
Eq. (\ref{Jmn}) should be retained. If $n+m=2P+1$ is odd, then $J$ is an odd
function of $y$. In this case the principal value integral gives a result of
order $z^{-1}$ and only the $L=1$ term is needed. 
The $\phi $ integral may then be performed, leaving
\begin{eqnarray}
I_{n+m=2P}=\frac{(2P)!}{2^{2P}(P!)^{2}}\int  &&dy{\mathcal{P}}\frac{%
z(1-y^{2})^{P}}{z^{-1}-y}  \label{even} \\
I_{n+m=(2P+1)}=-\frac{(2P+1)!}{2^{2P}(P!)^{2}}\int  &&dy(1-y^{2})^{P}.
\label{odd}
\end{eqnarray}
Finally, performing the $y$ integrals leads to
\begin{equation}
I_{nm}=(-1)^{n+m}.
\end{equation}
Substituting this into Eq.(\ref{phi13}), we find
\begin{equation}
\Phi _{3}^{(1,1)}=-\frac{\pi }{4}~\left( \sum_{n}(-1)^{n}{\tilde{\Gamma}}%
_{n}\right) ^{2}=-\frac{\pi }{4}\Gamma _{\text{BS}}^{2}.  \label{answer3d}
\end{equation}
We see that the contribution $\Phi _{3}^{(1,1)}$ is expressed solely in
terms of the backscattering amplitude, i.e., it comes exclusively from 1D
scattering processes. Notice that Eq. (\ref{answer3d}) can be also expressed
in terms of the angular harmonics of $\Gamma ^{k}$, introduced in Eq.(\ref
{leg}), as
\begin{equation}
\Phi _{3}^{(1,1)}=-\frac{\pi }{4}\Gamma _{\text{BS}}^{2}=-\frac{\pi }{4}%
~\left( \sum_{n}(-1)^{n}{\bar{\Gamma}}_{n}\right) ^{2}.  \label{answer3d_1}
\end{equation}

We pause here to emphasize the non-triviality of this result. As we have
already mentioned, the same process with one excited particle-hole pair
leads to a nonanalyticity in $D=2$.
For two dimensions,
however, an simple geometric argument, which we displayed in Section II,
shows that 1D processes determine the nonanalytic part of the entropy.
The straightforward extension of this argument to $D >2$ suggests that
 all relative angles are important, but Eq.(\ref{answer3d}) shows that $\Phi
_{3}^{(1,1)}$ still comes only from 1D scattering.

We next consider the contribution involving $\Gamma ^{(2)}$. Following the
arguments given above we find
\begin{equation}
\Phi _{3}^{(1,2)}=-\frac{\pi }{4}~\sum_{n,m,l}{\tilde{\Gamma}}_{n}{\tilde{%
\Gamma}}_{m}{\tilde{\Gamma}}_{l}I_{nml}.  \label{phi311}
\end{equation}
 Calculations along the same lines as above yield
\begin{equation}
I_{nm,l=(2P+1)}=\frac{2P+1!}{2^{2P}(P!)^{2}}I\left( \frac{n}{2}+p\right)
I\left( \frac{m}{2}+p\right)   \label{inml}
\end{equation}
with
\begin{equation}
I(m)=\int_{-1}^{1}\frac{dy}{2}(1-y^{2})^{m}.  \label{i_m}
\end{equation}
The integrals may be performed in terms of $\Gamma $-functions but the final
expressions seem complicated and not very useful in general. For further
comparison with Ref.\cite{Pethick73a}, we present here the result for $\Phi
_{3}^{(1,2)},$ obtained by truncating the sum in Eq.(\ref{phi311}) at $%
n,m,l=4$ and writing the result  in terms of the angular harmonics of $\Gamma ^{k}$:
\begin{eqnarray}
&&\Phi _{3}^{(1,2)}=-\frac{\pi }{4}~\left[ {\bar{\Gamma}}_{1}{\bar{\Gamma}}%
_{0}^{2}+\frac{2}{9}{\bar{\Gamma}}_{2}\left( 2{\bar{\Gamma}}_{0}^{2}+{\bar{%
\Gamma}}_{0}{\bar{\Gamma}}_{2}+\frac{4}{5}{\bar{\Gamma}}_{2}^{2}\right) %
\right]   \notag \\
&&-\frac{\pi }{12}\left[ {\bar{\Gamma}}_{1}^{2}{\bar{\Gamma}}_{2}+\frac{2}{45%
}{\bar{\Gamma}}_{4}\left( 4{\bar{\Gamma}}_{1}^{2}+3{\bar{\Gamma}}_{1}{\bar{%
\Gamma}}_{3}+\frac{18}{7}{\bar{\Gamma}}_{3}^{2}\right) \right] .
\label{phi311_1}
\end{eqnarray}

\subsubsection{Three  particle-hole pairs}

The second contribution to the nonanalytic term in the
specific heat of a three-dimensional Fermi Liquid comes from terms in Eq. (%
\ref{S3dfinal}) involving three excited particle-hole pairs (
e.g. diagram \textit{%
3c} in Fig. \ref{Fig:Omegas}).
This contribution gives the
``paramagnon'' term in the specific heat studied, e.g., in Ref.
\cite{Doniach66}. In this case the quasiparticle interaction
$\Lambda $ involves three factors of the static scattering
amplitude $\Gamma ^{k}$  whereas Im$\mathcal{P}$ can be taken in
the $z\rightarrow \infty $ limit, yielding

\begin{eqnarray}
&&\Phi _{3}^{(3)} =\frac{1}{192}\int d\mathbf{n}_{p_{1}}d\mathbf{n}_{p_{2}}d%
\mathbf{n}_{p_{3}}\delta (\mathbf{n}_{p_{1}}\cdot \mathbf{n}_{q})\delta (%
\mathbf{n}_{p_{2}}\cdot \mathbf{n}_{q})  \notag \\
&&\times\delta (\mathbf{n}_{p_{3}}\cdot
\mathbf{n}_{q})~ \Gamma ^{k}(\mathbf{n}_{p_{1}}\cdot \mathbf{n}_{p_{2}})\Gamma ^{k}(\mathbf{%
n}_{p_{2}}\cdot \mathbf{n}_{p_{3}})\Gamma ^{k}(\mathbf{n}_{p_{3}}\cdot
\mathbf{n}_{p_{1}}). \notag
\end{eqnarray}
Substituting Eq. (\ref{gamma3d}) into the equation for 
$\Phi _{3}^{(3)}$ and evaluating the integrals, we obtain
\begin{equation}
\Phi _{3}^{(3)}=\frac{\pi ^{3}}{48}~\sum_{lmn}{\tilde{\Gamma}}_{l}{\tilde{%
\Gamma}}_{m}{\tilde{\Gamma}}_{n}\frac{(l+m)!(l+n)!}{2^{n+m+2l}\left( \frac{%
l+m}{2}\right) !\left( \frac{n+m}{2}\right) !}.
\end{equation}
Truncating the sum at $l,m,n=2$ and re-expressing the result in terms of $%
\bar{\Gamma}_{L}$ [Eq.(\ref{leg})], we obtain
\begin{equation}
\Phi _{3}^{(3)}=\frac{\pi ^{3}}{48}~\left[ \left( {\bar{\Gamma}}_{0}+\frac{1%
}{4}{\bar{\Gamma}}_{2}\right) ^{3}+\frac{1}{4}{\bar{\Gamma}}_{1}^{3}+\frac{27%
}{256}{\bar{\Gamma}}_{2}^{3}\right]\label{aa}
\end{equation}

\subsubsection{The final result for the nonanalytic part of the entropy in 3D}

Combining all three expressions $\Phi_3^{(1,1)},~\Phi_3^{(1,2)}$ and 
$\Phi_3^{(3)}$,  and restoring the sum of charge and spin
components of $\Gamma ^{k}$, we finally obtain for the entropy per particle
\begin{eqnarray}
&&\delta S_{\text{NA}}(T)=-\frac{\pi ^{4}}{5} \left( \frac{T}{v_{F}^{\ast }k_{F}}%
\right) ^{3}\ln \left( \frac{v_{F}^{\ast }k_{F}}{T}\right) (K_{c}+3K_{s}),
\notag \\
&&K_{a}=\Gamma _{a,\text{BS}}^{2}+{\bar{\Gamma}}_{a,1}{\bar{\Gamma}}%
_{a,0}^{2}+\frac{1}{3}{\bar{\Gamma}}_{a,1}^{2}{\bar{\Gamma}}_{a,2}  \notag \\
- &&\frac{\pi ^{2}}{12}\left( \left( {\bar{\Gamma}}_{a,0}+\frac{1}{4}{\bar{%
\Gamma}}_{a,2}\right) ^{3}+\frac{1}{4}{\bar{\Gamma}}_{a,1}^{3}+\frac{27}{256}%
{\bar{\Gamma}}_{a,2}^{3}\right) +...  \label{S3dfinal_1}
\end{eqnarray}
 where  $a=c,s,$ $\Gamma _{a,%
\text{BS}}=\Gamma _{a}(\infty ;\mathbf{n}_{k},-\mathbf{n}_{k}),$ and dots
stand for the terms involving ${\bar{\Gamma}}_{a,n}$ with $n>2$.

A comment is in order here. In the consideration above we assumed that
 the backscattering amplitudes $\Gamma_{a,\text{BS}}$ are
 temperature-independent. Strictly speaking, this is not the case as the backscattering
amplitudes describe processes with total zero momentum and thus
can be re-expressed via the partial components of the pairing
vertex~\cite{aleiner_1}. One of these partial components diverges
at the pairing instability which, for a repulsive interaction, is
of the Kohn-Luttinger type~\cite{kohn_luttinger}), so
 our consideration is only valid at $T > T_c$.  Even at these temperatures,
the backscattering amplitudes acquire a logarithmic
temperature dependences from  the Cooper channel, and behave as
\begin{equation}
\Gamma _{a,\text{BS}}(T) \propto \sum_{n=0}^\infty
\frac{1}{\left(\alpha_n+\beta_n\ln T/E_F\right)}, \label{aug21_1}
\end{equation}
where $\alpha_n$ and $\beta_n$ are constants.
At $T_c$, $\alpha_{n} + \beta_n\ln T/E_F =0$ for a particular $n = n_0$.
In the rest of the text, we neglect this complication and assume that
the system is substantially far away from $T_c$ that both spin and charge components of the
backscattering amplitudes can be approximated by constants.

\subsection{Ring Diagram (RPA) Approximation}

In this subsection we present results for a toy model in which the expansion
in the number of excited particle-hole pairs can be carried to all orders, in order
to  demonstrate explicitly that only the squares and the cubes
 of the partial components of the scattering amplitude, but not higher powers,
 determine  the non-analyticity in  $S(T)$.

We consider an artificial model of fermions with spin degeneracy $%
N\rightarrow \infty $, coupled by a contact interaction $U(\mathbf{n}%
_{p}\cdot \mathbf{n}_{k})=U_{0}+U_{1}\mathbf{n}_{p}\cdot \mathbf{n}_{k}$.
Standard techniques yield, for the large-N limit of the entropy $S_{\infty
}=-\lim_{N\rightarrow \infty }(d \Xi(N)/dT)/N$
\begin{eqnarray}
S_{\infty } &=&S_{F}+\frac{S_{ring}}{N} \\
S_{\text{ring}} &=&- \text{Im}\int \frac{d\varepsilon }{\pi }\frac{\varepsilon }{4T^{2}\sinh%
\frac{\varepsilon }{2T}}\int \frac{d^{3}q}{(2\pi )^{3}}  \notag \\
&\times &{\text{T}r}\ln \left[ \left( 1-\Gamma (\mathbf{n}_{p}\cdot \mathbf{n%
}_{k})\mathcal{P}(z,\mathbf{n}_{k})\right) \right],  \label{SN}
\end{eqnarray}
where $S_{F}=-
d/dT \left[\text{Tr}\ln G_{0}^{-1}\right]$ the free fermion entropy,
 the trace taken over
angle ($\text{Tr}=\prod d n_{k}$), $\Gamma =U(\mathbf{n}_{p}\cdot \mathbf{n}%
_{k})k_{F}^{2}/(\pi ^{2}v_{F}^{\ast })=\Gamma _{0}+\Gamma _{1}\mathbf{n}%
_{p}\cdot \mathbf{n}_{k}$ and

\begin{equation}
{ \mathcal P}_{ph}(z,\mathbf{n}_{k})=-\frac{1}{4\pi }\frac{\cos\left( \mathbf{n}%
_{k}\cdot \mathbf{n}_{q}\right) }{\cos\left( \mathbf{n}_{k}\cdot \mathbf{n}%
_{q}\right) -z^{-1}}.
\end{equation}
Note that  ${\mathcal P}_{ph}$ is the full propagator; we do not
split it into analytic and nonanalytic parts. Making the changes
of variables to $x = \varepsilon/(2T)$
 and $z = v^*_F q/\varepsilon$, and integrating over $x$
as before we obtain for the entropy per particle
\begin{eqnarray}
S_{ring} &=&- \frac{2\pi ^{3}}{5}\left( \frac{T}{v_{F}^{\ast }k_{F}}\right)
^{3}\int_{1}^{v/Tx}z^{2}dz  \notag \\
&\times &\text{Im}\ln \text{Det}\left[ 1-\Gamma
(\mathbf{n}_{p}\cdot
\mathbf{n}_{k})\mathcal{P}(z,\mathbf{n}_{k})\right].  \notag
\label{SN2}
\end{eqnarray}

To evaluate the determinant we must find the eigenvalues of the operator $%
1-\Gamma \mathcal{P}$, i.e. we must solve
\begin{equation}
\lambda \Lambda (z,\mathbf{n}_{p})=\Lambda (z,\mathbf{n}_{p})-\int d\mathbf{n%
}_{k}\Gamma \left( \mathbf{n}_{p}\cdot \mathbf{n}_{k}\right)
{\mathcal P}_{ph}(z,%
\mathbf{n}_{k})\Lambda (z,\mathbf{n}_{k}).
\label{ev3d}
\end{equation}
Writing
\begin{equation}
\Lambda(\mathbf{n}_p)=\Lambda_0+\Lambda_1\cos\theta_p+\Lambda_2
\sin\theta_p(e^{i\phi}+e^{-i\phi})  \label{lamansatz}
\end{equation}
and using Eq.(\ref{gamma3d}), we find that the determinant may be
written as the product $\text{Det}=D_1 \times D^2_2$ ($\text{Im}
\ln \text{Det} = \text{Im} \ln D_1 + 2 \text{Im} \ln D_2$), with
\begin{eqnarray}
D_1&=&\left(1+\Gamma_0I_{00}\right)\left(1 +\Gamma_1I_{11}\right)
-\Gamma_0\Gamma_1I_{01}^2  \label{D1} \\
D_2&=&1 + \Gamma_1{\bar I}_{11}  \label{D2}
\end{eqnarray}
and
\begin{eqnarray}
I_{nm}&=&\int_{-1}^1\frac{d (\cos\theta)}{2}\frac{(\cos\theta)^{n+m+1}}{%
\cos\theta-z^{-1}}  \label{Inm} \\
{\bar I}_{11}&=&\int_{-1}^1\frac{d (\cos\theta)}{4}\frac{\cos\theta(\sin\theta)^2%
}{\cos\theta-z^{-1}}  \label{barI} 
\end{eqnarray}

Evaluating $I_{mn}$, we find
\begin{eqnarray}
I_{00}&=&1+\frac{1}{2z}ln\frac{1-z^{-1}}{1+z^{-1}}+i\frac{\pi}{2z} \\
I_{01}&=&\frac{1}{z}I_{00} \\
I_{11}&=&\frac{1}{3}+\frac{1}{z^2}I_{00} \\
{\bar I}_{11}&=& \left(I_{00}-I_{11}\right).
\label{xxxx}
\end{eqnarray}
Substituting Eqs.(\ref{xxxx}) into Eqs.(\ref{D1},\ref{D2})
 and simplifying yields
\begin{eqnarray}
D_1&=& A \left(1+{\bar \Gamma}_0(I_{00}-1)\right)\left(1+{\bar \Gamma}%
_1\left(I_{11}-1/3)\right)\right)  \notag \\
&&-{\bar \Gamma}_0{\bar \Gamma}_1I_{01}^2 \\
D_2&=&B +{\bar \Gamma}_1({\bar I}_{11}-1/3)
\end{eqnarray}
with $A= (1+\Gamma_0) (1 +\Gamma_1/3),~B = 1+\Gamma_1/3$, and the reducible
amplitudes ${\bar \Gamma}$ given by
\begin{eqnarray}
{\bar \Gamma}_0=\frac{\Gamma_0}{1+\Gamma_0} \\
{\bar \Gamma}_1=\frac{\Gamma_1}{1+\frac{\Gamma_1}{3}}.
\end{eqnarray}

The $z$ integral in Eq [\ref{SN2}] is dominated by its upper limit. The
leading term is $\mathcal{O}(z^{-1})$ and gives a renormalization of the
effective mass. The next term is $\mathcal{O}(z^{-3})$ and gives the
logarithm we require. Expanding
in $1/z$, we obtain
\begin{eqnarray}
\text{Im}\ln D_1 &\rightarrow & \frac{i\pi {\bar{\Gamma}}_{0}}{2z} +\frac{i\pi {%
\bar{\Gamma}}_{1}}{2 z^{3}} +\frac{i\pi \left( {\bar{\Gamma}}_{0}^{2}-2{\bar{%
\Gamma}}_{0}{\bar{\Gamma}}_{1}\right) }{2z^{3}}  \notag \\
&&-\frac{i\pi ^{3}{\bar{\Gamma}}_{0}^{3}}{24z^{3}}+ \frac{i\pi {\bar{\Gamma}}%
_{0}^{2}{\bar{\Gamma}}_{1}}{2z^{3}} \label{p_1}\\
\text{Im} \ln D_2 &\rightarrow&\frac{i\pi {\bar{\Gamma}}_{1}}{4z}-\frac{i\pi {\bar{\Gamma}}%
_{1}}{4z^{3}}+\frac{i\pi {\bar{\Gamma}}_{1}^{2}}{4z^{2}}-\frac{i\pi {\bar{%
\Gamma}}_{1}^{3}}{192z^{3}}.
\label{p_2}
\end{eqnarray}
Combining (\ref{p_1}) and (\ref{p_2}) to obtain
 $\text{Im} \ln \text{Det}=\text{Im} \ln D_1 + 2 \text{Im} \ln D_2$, extracting the $%
z^{-3}$ term and performing the integral in Eq.(\ref{SN2}) yields for
entropy per particle
\begin{eqnarray}
 S_{ring} &=&-\frac{\pi ^{4}}{5}\left( \frac{T}{v^*_F k_F}\right)^{3}
\ln\frac{v k_{F}}{T} \\
&&\left( \left( {\bar \Gamma} _{0}-{\bar \Gamma}_{1}\right)^{2}+{\bar \Gamma}
_{0}^{2}{\bar \Gamma} _{1}-\frac{\pi ^{2}{\bar\Gamma} _{0}^{3}}{12} -\frac{%
\pi ^{2}{\bar\Gamma} _{1}^{3}}{48}\right)  \notag  \label{Sringfinal}
\end{eqnarray}
This expression is in agreement with our previous result, Eq (\ref
{S3dfinal_1}) once we note that to the order in $N$ to which
we work, only
the charge component of $%
\Gamma$ contributes to a nonanalytic $\delta S (T)$, and that in the RPA approximation, the
dimensionless $\Gamma$ coincides with the Landau interaction function (this
also implies that ${\bar \Gamma}_n$ are the partial components of the
scattering amplitude). Note that, as before, the terms with coefficient $\pi
^{2}$ in (\ref{Sringfinal}) arise from processes with
three  excited particle-hole pairs while the others involve
processes with one excited particle-hole pair, along with a non-analyticity
in the quasiparticle interaction.

\section{Non-analyticity in Specific Heat in Dimension D=2}

\subsection{Overview}

For completeness, we here present results for the nonanalytic contribution $%
\delta S\propto T^{2}$ to the entropy a two- dimensional Fermi Liquid. This
contribution was found in Ref. \cite{Chubukov04,Chubukov05} by exact evaluation
of diagrams up to third order and by showing that results obtained by
analysis of leading singular behavior of diagrams occurring in the general
case matched the perturbative results exactly. The calculations
were also performed in  Matsubara frequencies.
Here we show how the same result is obtained using an expansion in the  number
of real particle-hole pairs. In addition, we reconfirm the result that the
non-analyticity in 2D involves only backscattering processes. Unlike the 3D
case, this is the only non-analytic contribution to $S(T)$.

We begin from the two dimensional version of Eq [\ref{SNA}] in which the essential
object is $\int_1^{v^*_Fk_F/Tx} zdz\Phi_{2D}(z)$. Again we expect the leading
large $z$ behavior of $\Phi$ to be $\Phi_{2D}\sim 1/z$ giving the Fermi-liquid behavior;
but the different ultraviolet behavior of the momentum integral
means that the remaining terms are ultraviolet
convergent, so that the $z$ integral is just a constant and the temperature prefactor
gives the nonanalytic result in 2D: $\gamma=d(\delta S_{NA})/dT \sim T$.

A further important question arises from the
analytic structure of $\Pi $. In $D=2,$
\begin{eqnarray}
\text{Re}\Pi (z) &=&\frac{\Theta \left( 1-\left| z\right| \right) }{\sqrt{%
1-z^{2}}},  \label{re2} \\
\text{Im}\Pi (z) &=&-\frac{\text{sgn}z\Theta \left( \left| z\right|
-1\right) }{\sqrt{z^{2}-1}}.  \label{im2}
\end{eqnarray}
The singularity at the boundary of the particle-hole continuum $|z|=1$ has
be treated with care when expanding $\delta S_{\text{NA }}$in powers
of Im$\Pi $, and indeed raises the possibility that the $z$ integral
has appreciable contributions from the neighborhood of  $z=1$, where 
the kinematics of Landau damping does not guarantee that the contributions come from fermions
with momentum perpendicular to $q$, and therefore might not be determined
solely  by the backscattering amplitude.

The computation is in fact most straightforwardly carried out by rotating
the $\varepsilon $ integral in Eq. (\ref{S}), leading in $D=2$ to the
non-analytic entropy per particle
\begin{eqnarray}
\delta S_{\text{N}A} &=&-\frac{d\Xi _{\text{NA}}(T)}{dT}~,  \notag \\
~~\Xi _{\text{NA}}(T) &=&\left( \frac{1}{2v_{F}^{\ast }k_{F}}\right)
^{2}T\sum_{\varepsilon _{m}}\varepsilon _{m}^{2}\int_{0}^{\frac{v_{F}^{\ast
}k_{F}^{{}}}{|\varepsilon _{m}|}}dx~x\Psi (ix), \nonumber \\
 \label{S2d}
\end{eqnarray}
where $x=v_F^{\ast }q/i|\varepsilon _{m}|$ and
\begin{equation}
\Psi (z)=\int \frac{dy}{\pi }\frac{\Phi (y)}{z-y}
\end{equation}
is the function whose discontinuity across the real axis is equal to $\Phi $
from Eq. (\ref{S}).

Nonanalytic contributions to Eq.(\ref{S2d}) can only arise from the
dependence of the integral over $x$ on its upper limit, which is determined
by the behavior of $\Psi $ at large argument. Writing
\begin{equation}
\Psi (z\rightarrow \infty )\rightarrow \frac{\Psi _{1}}{z^{{}}}+\frac{\Psi
_{2}}{z^{2}}+...
\end{equation}
yields
\begin{equation}
\Xi _{\text{N}A}(T)=\frac{\Psi _{2}}{4(v_{F}^{\ast }k_{F})^{2}}%
~T\sum_{\varepsilon _{m}}\varepsilon _{m}^{2}\ln \frac{v_{F}^{\ast }k_{F}}{%
|\varepsilon _{m}|}.  \label{S2d_1}
\end{equation}
The frequency sum generates regular terms determined by the upper limit of
the summation; these form an analytic expansion of $\Xi (T)$ in powers of $%
T^{2}$. However, because of the logarithm, the sum also produces a
universal, i.e., independent of the upper limit, term of order $T^{3}$
\begin{equation}
T\sum_{\varepsilon _{m}}\varepsilon _{m}^{2}\ln \frac{v_{F}^{\ast }k_{F}}{%
|\varepsilon _{m}|}=-2\zeta (3)T^{3}+...  \label{S2d_2}
\end{equation}
where $\dots $ stand for regular terms. This universal term can be obtained
either directly, by using Euler-Maclaurin summation formula, or by
evaluating the frequency sum as a contour integral. Substituting (\ref{S2d_2}%
) into (\ref{S2d_1}) and differentiating over $T$, we obtain
\begin{equation}
\delta S_{\text{NA}}=\frac{3\zeta (3)}{2}~\Psi _{2}~\left(
\frac{T}{v_{F}^{\ast }k_{F}}\right) ^{2}+\mathcal{O}(T^{3}).
\label{S2d2}
\end{equation}

\subsection{Expansion in Particle-Hole Pairs}

First, we observe that because each factor of $\mathcal{P}_{\text{NA}}$
brings an extra factor of $z^{-1}$, the only contribution to $\Psi _{2}$
involves one real excited particle-hole pair and the leading correction to $%
\Gamma $, and is thus proportional to the square of the reducible
interaction. Proceeding as in the previous section, we obtain  the
analogue of Eq. (\ref{phi31})
\begin{equation}
\Psi _{2}^{(1)}=-\int d{\theta }_{p}\mathcal{P}(ix;{\theta }_{p})\Gamma (ix;{%
\theta }_{p},{\theta }_{p})  \label{phi21}
\end{equation}
and the analogue of Eq. (\ref{gam1})
\begin{equation}
\Gamma =\int d{\theta }_{k}\Gamma ^{k}({\theta }_{p}-\theta _{k})\text{Re}%
\mathcal{P}(z,\theta _{k})\Gamma ^{k}(\theta _{k}-\theta _{p}).
\label{gamma2d}
\end{equation}
Writing
\begin{equation}
\Gamma ^{k}({\theta }_{p}-\theta _{k})=\sum_{L}{\bar{\Gamma}}_{L}\cos \left[
L({\theta }_{p}-\theta _{k})\right] ,  \label{gammak2d}
\end{equation}
using $(\cos \theta -ix)^{-1}=-i\int d\lambda \exp [-i\lambda (\cos \theta
-ix)]$ (for $x>0$), expanding the exponential in terms of Bessel functions
and using known integrals gives
\begin{eqnarray}
&&\Psi^{1}_{2} (ix)=-\sum_{LL^{\prime }}\frac{{\bar{\Gamma}}_{L}{\bar{\Gamma}}%
_{L^{\prime }}}{2x^{2}\sqrt{1+x^{-2}}} \\
&&\times \left[ \left( \frac{1}{x}-\sqrt{1+x^{-2}}\right) ^{L+L^{\prime
}}+\left( \frac{1}{x}-\sqrt{1+x^{-2}}\right) ^{L-L^{\prime }}\right] ^{2}.
\notag  \label{2d}
\end{eqnarray}
Evidently, the large-argument limit of Eq.(\ref{2d}) is simply
\begin{equation}
\Psi _{2}=-\sum_{L,L^{\prime }}(-1)^{L+L^{\prime }}{\bar{\Gamma}}_{L}{\bar{%
\Gamma}}_{L^{\prime }}=-\left( \sum_{L}(-1)^{L}{\bar{\Gamma}}_{L}\right) ^{2}.
\end{equation}
We see that the nonanalytic term in the entropy is determined by the
backscattering amplitude. Restoring the spin and charge components of $%
\Gamma $, we then obtain for the entropy per particle
\begin{equation}
\delta S_{NA}=-\frac{3\zeta (3)}{2}\left( \frac{T}{v_{F}^{\ast }k_{F}^{{}}}\right)
^{2}~\left( \Gamma _{c,\text{BS}}^{2}+3\Gamma _{s,\text{BS}}^{2}\right) +%
\mathcal{O}(T^{3}).  \label{S2d2_1}
\end{equation}
This coincides with the result in Ref.\cite{Chubukov04}.

\subsubsection{Computation in real frequencies}

For the sake of completeness, we also demonstrate how to obtain Eq. (\ref
{S2d2}) by performing a computation directly in real frequencies, without rotating
the $\varepsilon $ integral in Eq.(\ref{S}). Using Eq. (\ref{SNA}) and the
two-dimensional analogues of Eqs. (\ref{phi31}) and (\ref{gam1}), we obtain
in 2D for the entropy per particle
\begin{widetext}
\begin{equation}
\delta S_{\text{NA}} = \frac{4}{\pi}~
 \left(\frac{T}{v^*_F k^{}_F}\right)^2
\int_{0}^{\infty }dx\frac{x^{3}}{\sinh ^{2}x}\int_{0}^{v_F^{\ast
}k_{F}/xT} dz z \int d\mathbf{n}_{\mathbf{k}}\int
d\mathbf{n}_{\mathbf{p}}~\text{Im} \Pi \left(
z,\mathbf{n}_{\mathbf{k}}\right)
\text{Re} \Pi \left( z,\mathbf{n}_{\mathbf{p}}\right)
\left[
\Gamma ^{k}\left( \mathbf{n}_{\mathbf{k}}\cdot
\mathbf{n}_{\mathbf{p}%
}\right) \right] ^{2}.
\label{ch1}
\end{equation}
\end{widetext}
We  know from Eq. (\ref{S2d2}) that $\delta S_{\text{NA }}\propto T^{2}$. As
$T^{2}$ is  an overall factor in Eq.(\ref{ch1}), the rest of the
integral can can be evaluated at $T=0$ upon which the upper limit of $z$
integral is set to infinity. Integrations over $x$ and over $z$ then
decouple, and evaluating the integral we obtain
\begin{equation}
\delta S_{\text{NA}}=\frac{6\zeta \left( 3\right) }{\pi }~\left( \frac{T}{%
v_{F}^{\ast }k_{F}}\right) ^{2}~R,  \label{ch7}
\end{equation}
where
\begin{eqnarray}
R &=&~\int_{0}^{\infty }dzz\int d\mathbf{n}_{\mathbf{k}}\int d\mathbf{n}_{%
\mathbf{p}}~\left[ \Gamma ^{k}\left( \mathbf{n}_{\mathbf{k}}\cdot \mathbf{n}%
_{\mathbf{p}}\right) \right] ^{2}  \notag \\
&&\times \text{Im}\mathcal{P}\left( z,\mathbf{n}_{\mathbf{k}}\right) \text{Re%
}\mathcal{P}\left( z,\mathbf{n}_{\mathbf{p}}\right) .  \label{ch2}
\end{eqnarray}
The subtlety, which we mentioned in the overview of this section, becomes
clear if we assume momentarily that $\Gamma ^{k}$ a constant. In this case,
the angular integrations in Eq.(\ref{ch2}) are performed independently, and $%
R$ reduces to $\left[ \Gamma ^{k}\right] ^{2}\int_{0}^{\infty }dzz$Im$\Pi
\left( z\right) $Re$\Pi \left( z\right) $. As Eqs.(\ref{re2},\ref{im2})
show, Im$\Pi \left( z\right) $ and Re$\Pi \left( z\right) $ are finite on
the opposite sides of the boundary of the the particle-hole continuum (for $%
\left| z\right| >1$ and $\left| z\right| <1,$ respectively).
One might then  conclude that
 $\delta S_{\text{NA}}=0.$ However,
the confluent singularities in $Re \Pi$ and $Im \Pi$ require a more careful analysis.
Retaining the infinitesimal $i\delta $ to $\Pi $  at  intermediate steps one finds
\begin{eqnarray}
&&\int_0^\infty dz z \text {Im} \Pi \left( z\right) \text{ Re}
\Pi \left( z\right)  \nonumber \\
&& =\int_0^\infty \frac{dz z}{z+1}~ \text {Im} \frac{1}{\sqrt{z-1
+ i \delta}} \text {Re}  \frac{1}{\sqrt{z-1 + i \delta}} \nonumber \\
&& =\frac{1}{2} \int_{-\infty}^\infty
dx  \text {Im} \frac{1}{\sqrt{x + i}}  \text {Re} \frac{1}{\sqrt{x
+ i}}\nonumber \\
&& = -\frac{1}{4} \int_{-\infty}^\infty \frac{dy}{y^2+1} =
-\frac{\pi}{4}.
\end{eqnarray}

With this in mind, we proceed with the case of an arbitrary vertex $\Gamma
^{k}\left( \mathbf{n}_{\mathbf{k}}\cdot \mathbf{n}_{\mathbf{p}}\right) $. As
before, we expand $\Gamma ^{k}$ in harmonics ${\bar{\Gamma}}_{L}$, see Eq.
(\ref{gammak2d}). Substituting this expansion into Eq.(\ref{ch2}), we obtain
after simple algebra
\begin{eqnarray}
R &=&
 2 \int_{0}^{\infty }dzz\int_{0}^{\pi }d\theta _{%
\mathbf{k}}\int_{0}^{\pi }d\theta _{\mathbf{p}}  \label{ch4} \\
&&\times \text{Im}\mathcal{P}\left( z,\theta _{\mathbf{k}}\right) \text{Re}%
\mathcal{P}\left( z,\theta _{\mathbf{p}}\right)  \notag \\
&&\times \sum_{L,L^{\prime }}{\bar{\Gamma}}_{L}{\bar{\Gamma}}_{L^{\prime }}%
\left[ \cos \left( L_{+}\theta _{\mathbf{k}}\right) \cos \left( L_{+}\theta
_{\mathbf{p}}\right) +\left( L_{+}\rightarrow L_{-}\right) \right] ,  \notag
\\
&=&\sum_{L,L^{\prime }}{\bar{\Gamma}}_{L}{\bar{\Gamma}}_{L^{\prime
}}I_{L,L^{\prime }}. \notag
\end{eqnarray}
where $L_{\pm }=L\pm L^{\prime }.$ As $R$ turns out to depend only on whether
$L_{\pm },$ is odd or even we can just consider the $L_{+}$ term and double the result.

Consider first contributions with odd $L_{+}$. For these contributions, as
we will see, one can neglect the $i\delta $ term in $\mathcal{P}$, i.e.,
replace Im$\mathcal{P}\left( z,\theta _{\mathbf{k}}\right) $ by a $\delta -$%
function 
Im$\mathcal{P}\left( z,\theta _{\mathbf{k}}\right) =
-(1/2z) \delta \left( z^{-1}-\cos \theta _{\mathbf{k}}\right) $. Integrating
then over $z$, we obtain
\begin{equation}
I_{L+L^{\prime }=2P+1}=\frac{1}{\pi }\left[ \int_{0}^{\pi /2}d\theta _{%
\mathbf{k}}\frac{\cos \left[ (2P+1)\theta _{\mathbf{k}}\right] }{\cos \theta
_{\mathbf{k}}}\right] ^{2}.  \label{R1}
\end{equation}
The integral is convergent and equal to $(-1)^{P}\pi /2$. Accordingly,
\begin{equation}
I_{L+L^{\prime }=2P+1}=\frac{\pi }{4}.  \label{ch5}
\end{equation}
If $L+L^{\prime }$ is even, the integral over $z$ vanishes unless one keeps $%
\delta $ finite and takes into account that both Re $\mathcal{P}$ and Im $%
\mathcal{P}$ diverge as $z$ approaches $1$. This divergences come from the
angular integral over a narrow range near $\theta _{k}=\theta _{p}=0$.
Accordingly, we can safely set $\theta _{k}=\theta _{p}$ outside the product
Re $\mathcal{P}$ $\times $Im $\mathcal{P}$ in Eq.(\ref{ch4}).
After simple algebra we then obtain
\begin{equation}
I_{L+L^{\prime }=2P}=-\frac{\pi }{4}.  \label{ch6}
\end{equation}
Combining Eqs.(\ref{ch5}) and (\ref{ch6}), we obtain
\begin{equation}
R=-\frac{\pi }{4}\sum_{L,L^{\prime }}(-1)^{L+L^{\prime }}{\bar{\Gamma}}_{L}{%
\bar{\Gamma}}_{L^{\prime }}=-\frac{\pi }{4}\left( \sum_{L}(-1)^{L}{\bar{%
\Gamma}}_{L}\right) ^{2}.  \label{ch5_1}
\end{equation}
Substituting this result into Eq. (\ref{ch7}) and introducing the charge and
 spin components of $\Gamma $, we reproduce Eq. (\ref{S2d2_1}).

\subsection{Ring Diagram (RPA) Approximation in 2D}

In this subsection we present the two-dimensional version of the
large-N calculation given above. Our goal here is to show that
only $\Gamma^2_{\text BS}$ determines the non-analyticity
in $S(T)$, whereas
higher powers of $\Gamma_L$ do not appear in the prefactor
for the $T^2$ term in $S(T)$.

We again consider a simplified model with the effective
interaction $U=U_{0}+U_{1}\cos(\phi _{p}-\phi _{k})$, and spin
degeneracy $N=\infty $. The two dimensional version of Eq
[\ref{SN2}] is
\begin{eqnarray}
&&S_{ring}=- \frac{6\zeta (3)}{\pi } \left(\frac{T}{v^*_F k_F}\right)^2~\times \nonumber \\
&& \int zdz{\text{Im}} \ln {\text{Det}}\left[ 1-\Gamma (\phi
_{p}-\phi _{k})\mathcal{P}(z,\phi _{k})\right],  \label{SN2d}
\end{eqnarray}
where now $\Gamma =Uk_{F}/(\pi v_{F}^{\ast })$ and
\begin{equation}
{\bar {\mathcal P}(z,\phi _{k})}=-\frac{1}{2\pi }\frac{\cos\phi
_{k}}{\cos\phi _{k}-z^{-1}}.  \label{Piz2d}
\end{equation}
To evaluate the determinant we again must find the eigenvalues of
the operator $1-\Gamma {\bar {\mathcal P}}$, i.e. solve
\begin{equation}
\lambda \Lambda (z,\mathbf{n}_{p})=\Lambda (z,\mathbf{n}_{p})-\int d\mathbf{n%
}_{k}\Gamma \left( \mathbf{n}_{p}\cdot \mathbf{n}_{k}\right) {\bar {\mathcal P}}(z,%
\mathbf{n}_{k})\Lambda (z,\mathbf{n}_{k}) \label{ev2d}
\end{equation}
We find the eigenfunctions of $1-\Gamma {\bar {\mathcal P}}$ by making the ansatz $%
\Lambda =\Lambda _{0}+\Lambda _{1}\cos\phi +\Lambda _{2}\sin\phi
$. We find that $\text{Det}=D_{1}D_{1}$ with $D_{1,2}$ given by
Eqs.(\ref{D1},\ref{D2}) as before, but now instead of
Eqs.(\ref{Inm},\ref{barI}) we have
\begin{subequations}
\begin{eqnarray}
I_{nm} &=&\int \frac{d\phi }{2\pi }\frac{(\cos\phi
)^{n+m+1}}{\cos\phi -z^{-1}}
\\
{\bar{I}}_{11} &=&\int \frac{d\phi }{2\pi }\frac{(\sin\phi )^{2}\cos\phi }{%
\cos\phi -z^{-1}}.
\end{eqnarray}
\end{subequations}
Explicitly
\begin{subequations}
\begin{eqnarray}
I_{00} &=&1+\frac{i}{\sqrt{z^{2}-1}} \\
I_{01} &=&\frac{1}{z}+\frac{i}{z\sqrt{z^{2}-1}} \\
I_{11} &=&\frac{1}{2}+\frac{1}{z^{2}}+\frac{i}{z^{2}\sqrt{z^{2}-1}} \\
{\bar{I}}_{11}
&=&\frac{1}{2}-\frac{1}{z^{2}}+\frac{i\sqrt{z^{2}-1}}{z^{2}}.
\end{eqnarray}
\end{subequations}
Evaluating and simplifying yields
\begin{subequations}
\begin{eqnarray}
D_{1} &=&\left(1 + \Gamma\right) \left(1 + \frac{\Gamma_1}{2}
\right) \times
\nonumber \\
&&\left(1+\frac{{\bar{\Gamma}}_{1}(1-{\bar{\Gamma}}_{0})}{z^{2}}+\frac{%
i\left( {\bar{\Gamma}}_{0}+\frac{{\bar{\Gamma}}_{1}(1-{\bar{\Gamma}}_{0})}{%
z^{2}}\right) }{\sqrt{z^{2}-1}}\right) \\
D_{2} &=& \left(1 + \frac{\Gamma_1}{2} \right) \times 
\left(1-\frac{{\bar{\Gamma}}_{1}}{z^{2}}+\frac{i{\bar{\Gamma}}_{1}\sqrt{%
z^{2}-1}}{z^{2}}\right)
\end{eqnarray}
\end{subequations}
with
\begin{subequations}
\begin{eqnarray}
{\bar{\Gamma}}_{0} &=&\frac{\Gamma _{0}}{1+\Gamma _{0}} \\
{\bar{\Gamma}}_{1} &=&\frac{\Gamma _{1}}{1+\frac{\Gamma _{1}}{2}}
\end{eqnarray}
\end{subequations}
Combining $\ln D_1$ and $\ln D_2$, we find
\begin{eqnarray}
&&\int_0^{v^*_F k_F/T} z dz \text{Im} \ln \text{Det}[1-{\bar
\Gamma} \mathcal{P}]=
\int_0^{v^*_F k_F/T} z dz \text{Im} \ln Q (z) \nonumber \\
&&Q (z) = 1-2 \frac{{\bar \Gamma} _{0}{\bar \Gamma} _{1}}{z^{2}}-%
\frac{2{\bar \Gamma} _{1}^{2}(1-{\bar \Gamma} _{0})}{z^{4}} -\nonumber \\
&&
\frac{1}{\sqrt{1- z^{2}}} \left({\bar \Gamma} _{0}+{\bar \Gamma} _{1}-\frac{%
2{\bar \Gamma} _{1}{\bar \Gamma} _{0}}{z^{2}}-
{\bar \Gamma} _{1}^{2}(1-{\bar \Gamma} _{0})\left( \frac{-1%
}{z^{2}}+\frac{2}{z^{4}}\right)\right). \nonumber \\
&& \label{xx_1}
\end{eqnarray}
The integral can be most straightforwardly evaluated by rotating
the $z-$integral onto the imaginary axis $z \rightarrow -i {\bar z}$.
Then
\begin{equation}
\int_0^{v^*_F k_F/T} z dz \text{Im} \ln Q (z) = -  \int_0^{i v^*_F
k_F/T} {\bar z} d{\bar z} \text{Im} \ln Q(-i {\bar z})
\label{xx_2}
\end{equation}
As $Q (-i {\bar z})$ is real and positive, the imaginary contribution from the
integral over $\bar z$  can only come from the upper limit. One
can verify that the dependence of the integral
in (\ref{xx_2}) on the upper limit  comes only from the
$O(1/{\bar z})$ and $O(1/{\bar z}^2)$ terms in the expansion of
$Im \ln Q(-i {\bar z})$ in powers of $1/{\bar z}$. Higher order
terms in the expansion are ultraviolet convergent and can be safely evaluated by
setting the upper limit to infinity. Expanding $ \ln Q(-i {\bar z})$ to
second order in $1/{\bar z}$, we obtain
\begin{equation}
 \ln Q(-i {\bar z}) =
  1 - \frac{{\bar \Gamma} _{0} +{\bar \Gamma} _{1}}
  {{\bar z}} - \frac{1}{2} \frac{({\bar \Gamma} _{0}- {\bar \Gamma} _{1})^2}{{\bar z}^{2}}
   + \mathcal{O} \left({\bar z}^{-3} \right).
\label{xx_3}
\end{equation}

As in 3D, the leading  term $O\left({\bar z}^{-1}\right)$ gives a
renormalization of the effective mass, while the next term
$O\left({\bar z}^{-2}\right)$ gives
\begin{eqnarray}
&&- \int_0^{i v^*_F k_F/T} {\bar z} d{\bar z} \ln Q(-i {\bar z}) \rightarrow \nonumber \\
&& \frac{1}{2} ({\bar \Gamma} _{0}- {\bar \Gamma} _{1})^2
\text{Im} \ln [i v^*_F k_F/T] =\frac{\pi}{4} ({\bar \Gamma} _{0}-
{\bar \Gamma} _{1})^2 \label{xx_4}
\end{eqnarray}
Substituting Eq.(\ref{xx_4}) into Eq.(\ref{SN2d}), we reproduce
Eq. (\ref{S2d2_1}).

\section{Spin susceptibility in 3D}

In this section we discuss the relation between nonanalyticities in the entropy
and in the spin susceptibility, $\chi_s$. Until recently, the 
prevailing opinion had been that
the nonanalytic $T^3 \ln T$ dependence of the entropy is not paralleled by
 a nonanalyticity in $\chi_s$. Crucial evidence for this view
was provided by the results of Carneiro and Pethick~\cite{Carneiro77} 
and B{\`e}al-Monod et al. \cite{Bealmonod}, who found that the leading 
term in the spin susceptibility scales as $T^2$ in 3D. However, in an
important paper  Belitz et al. \cite{belitz} demonstrated that
the apparent analytic temperature dependence of $\chi_s$ may be misleading.
They performed a perturbative calculation of the momentum-dependent spin
susceptibility $\chi_s (Q,T=0)$ at small $Q$ and
found a non-analytic $Q^2 \ln Q$ behavior.
Later, it was found \cite{betouras05} that the magnetic-field dependence
of a non-linear spin susceptibility parallels the $Q$-dependence, i.e.,
$\chi_s (Q=0, T=0, H) \propto H^2 \ln |H|$.

Nonanalyticity of the spin susceptibility was also found for 2D systems by
Millis and Chitov~\cite{chitov_millis} and, later, Chubukov and
Maslov~\cite{Chubukov04}, Galitski et
al.~\cite{galitski05}, and Betouras et al~\cite{betouras05}. 
 These authors showed
that $\chi_s (T, Q, H)$ scales linearly with the largest
out of the three parameters (in proper units).  Furthermore, 
Chubukov and Maslov~\cite{Chubukov04} and Galitski et
al.~\cite{galitski05} have shown that in 2D, the nonanalytic term in $\chi_s$
may be  expressed in terms of the spin component of the backscattering amplitude.

\begin{figure}[tbp]
\begin{center}
\epsfxsize=0.5\columnwidth
\epsffile{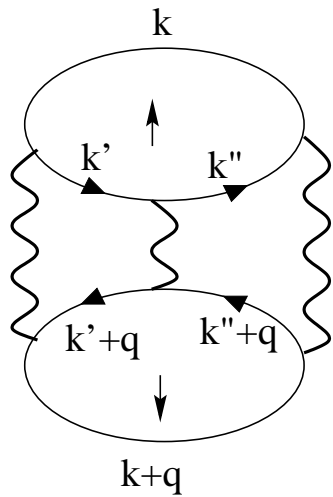}
\end{center}
\caption{A representative diagram for the thermodynamic potential, 
which contributes to 
the non-analyticity in the spin susceptibility to third order in the interaction. Vertical arrows denote  spin projections for fermions in the corresponding bubbles.}
\label{Fig:spin}
\end{figure}

These results call for a better understanding of the spin susceptibility of a
3D system. In the preceding sections, we found that
the nonanalytic part of the specific heat consists of
two contributions, the first one coming the excitation of a single particle-hole pair 
and the second one coming from the excitation of three 
particle-hole pairs, and with all coefficients determined by the 
 harmonics of the Landau
function. Here we ask if the non-analytic contributions to the spin susceptibility
are given in the same way. It is clear from the previous second-order calculations in 3D 
~\cite{belitz,chitov_millis,Chubukov04}
that the single-pair contribution does contribute to the non-analyticity in
$\chi_s (Q, T, H)$, as is shown by, e.g., the second-order 
calculations. We show here that the three-pair terms, which were not considered
in the previous literature, also contributes to the nonanalyticity
of $\chi_s$ in $3D$. For brevity, we refrain from a fully detailed
analysis and simply indicate the origin of the effect.
To this end, we consider the thermodynamic potential  
in a weak magnetic field  perturbatively to third order 
in the interaction, which we take to be momentum-independent.
 We will be only interested in  the spin effect of the field.
A magnetic field ($H$) splits the Fermi surfaces for fermions with
spins parallel and antiparallel to the field. This splitting does
not affect the $\varepsilon/q$ nonanalyticity of the polarization bubble
at small $q$, if a 
particle and a hole have the same spin  (in this case, a
magnetic field just shifts the chemical potential), but it has a
nontrivial effect on a bubble composed of a particle and a hole of
opposite spins, $\Pi_{\uparrow\downarrow}(\varepsilon,q;H)$. For a
spin-invariant interaction, such bubbles occur in diagram
Fig.~\ref{Fig:Omegas} \textit{3d}. Labeling the momenta as shown
in the figure and integrating over fermionic momenta ${\bf k}$,
${\bf k}'$, and ${\bf k}''$ and corresponding frequencies, we obtain 
the thermodynamic potential at $T=0$ and finite $H$ in the form
\begin{equation}
\Xi_{3D}(H)\sim U^3\int d\varepsilon\int d^3q
\text{sgn}\varepsilon\left[\text{Im}\Pi_{\uparrow\downarrow}(\varepsilon,q;H)\right]^3,
\label{XiH}
\end{equation} 
where
\[\text{Im}\Pi_{\uparrow\downarrow}(\varepsilon,q;H)=
-\left(\pi\varepsilon/2q\right)\Theta\left(v_Fq-|\epsilon-g_{L}\mu_B
H|\right)\]
and $g_L$ and $\mu_B$ are
 the Lande-factor and the Bohr magneton, respectively.
Note that $\text{Im}\Pi$ retains its $H=0$ form but the boundary of the 
continuum  is
shifted by the Zeeman energy. It is this shift which leads to a
nonanalyticity in $\Xi(H)$. Integrating over $q$ in Eq.(\ref{XiH})
yields
\begin{equation}
\Xi(H)\sim U^3\int
d\varepsilon|\varepsilon|^3\ln|\varepsilon-g_L\mu_BH|.
\label{XiH1}\end{equation} Consequently, the nonanalytic part of
$\Xi(H)$ behaves as $H^4\ln|H|$ and thus $\chi_s(H)\propto
H^2\ln|H|$.
This contribution has the same functional form as the contribution 
from the spin component of the
single-pair process. A similar consideration for a non-uniform
field leads to a nonanalytic $Q$ dependence of $\chi_s (Q)$: 
$\chi_s (Q) \propto Q^2 \ln Q$.

Finally, at vanishing $H$ and $Q$, both single-pair and three-pair
processes contribute  $T^2$ terms to the 3D
spin susceptibility, which again is expressed only in terms of the
Fermi-surface interaction and Fermi velocity.
However, it can hardly be identified in the experiment, as there  are additional, purely
analytic  $T^2$ contributions to $\chi_s (T)$ from all energies.

\section{Fermionic self-energy}

\subsection{Overview}

In this section we show how to obtain the non-analytic term in the entropy
and the specific heat by evaluating the fermionic self-energy first, and
then using the relation between the self-energy and thermodynamic potential.
We present this calculation both because historically the nonanalytic
terms in the specific heat were studied by first calculating the self energy
and because, as we will show, the self energy exhibits several nonanalyticities,
not all of which contribute to the specific heat.

It has been known since the early days of the Fermi-liquid theory
 that at $T=0$ the subleading term in the 3D fermionic  self-energy
on the Fermi surface $\Sigma (\omega,k_F)$
depends on the frequency in a nonanalytic way,
as  $\Sigma (\omega ,k_{F})\propto \omega ^{3}\ln \omega $.
Amit et al.~\cite{Amit68} and others argued that this $\omega ^{3}\ln \omega
$ dependence of $\Sigma $ is the source for the $T^{3}\ln T$ behavior of the
entropy and the specific heat, as the entropy has the same scaling dimension
as $\Sigma $ and typical frequencies $\omega $ are generally expected to be
of order $T$.
Indeed, to second order in the interaction $U$, the
thermodynamic potential in terms of the Matsubara self-energy is given by
\begin{equation}
\Xi =-\frac{1}{2}T\sum_{\varepsilon _{m}}\int \frac{d^{D}k}{(2\pi )^{D}}%
\Sigma (\omega _{m},k)G_{0}(\omega _{m},k)  \label{ch_10}
\end{equation}
where $G_{0}(\omega ,k)=(i\omega _{m}-\epsilon _{k})^{-1}$, and
$G^{-1}(\omega ,k) = G^{-1}_{0}(\omega ,k) + \Sigma (\omega, k)$. If the
self-energy is independent of $k$ (which is the case for, e.g., the
electron-phonon interaction), the momentum integral involves only the
Green's function, and $\int d^{D}kG_{0}(\omega _{m},k) \propto ${sgn}$\omega$.
 Therefore, $\Xi \propto T\sum_{\omega
_{m}>0}\Sigma (\omega _{m})$. Evaluating the frequency sum, one
obtains that the $\omega _{m}^{3}\ln |\omega _{m}|$ behavior of the
self-energy in 3D gives rise to a $T^{3}\ln T$ term in the entropy.

In general,  $\Sigma (\omega _{m},k)$ depends on both $\omega _{m}$
and $k.$ We will see that in the low-energy limit,
the self-energy in 3D is a sum of two terms, one depending
on $i\omega - \epsilon_k$, another depending separately on $\omega_m$ and $\epsilon_k$:
\begin{equation}
\Sigma (\omega _{m},k)=\Sigma ^{\left( 1\right) }\left( i\omega
_{m}-\epsilon _{k}\right) +\Sigma ^{\left( 2\right) }\left( \omega
_{m},\epsilon _{k}\right) .  \label{se12}
\end{equation}
On the Fermi surface ($\epsilon _{k}=0)$, the two terms
cannot be distinguished. However, away from the Fermi surface, their roles are
different. If only the first term is present then the full Green function
depends only on the variable $\omega-\varepsilon_k$ so that the locations
of the poles are not changed by the interaction and thermodynamics cannot
change.
Alternatively speaking,
the first term in Eq.(\ref{se12})  affects only the quasiparticle
renormalization factor $Z$, but as is well known, the $Z$ factor
does not affect the specific heat coefficient of a Fermi Liquid; hence the
first term cannot contribute to the non-analyticity in $S(T)$.

\subsection{Explicit calculation of the fermionic self-energy in 3D}

Now we proceed with an explicit evaluation of the nonanalytic part of
self-energy $\Sigma (\omega _{m},k)$ in 3D. To simplify the presentation, we
consider the self-energy only to second order in the interaction $U(q),$ and
 only discuss small-angle scattering, i.e., assume $U(0) =U,~U(2k_F) =0$
[cf. Eqs.(\ref{gammacfirstorder},\ref{gammasfirstorder})].

 As we mentioned
in Sec. II, the stationarity of $\Xi $ with respect to variations in $G$
implies \cite{LL,Pethick75} that one can neglect any explicit temperature
dependence of $\Sigma $, i.e., it suffices to include only the $T$-dependence
arising from the difference between $T\sum_{\omega _{m}}$ and $\int d\omega
/2\pi $.

The self-energy is given to leading order in $U$ by
\begin{equation}
\Sigma (\omega _{m},k)= U^{2}T\sum_{\varepsilon _{m}}\int \frac{d^{3}q}{%
(2\pi )^{3}}\Pi (\varepsilon _{m},q)G_{0}(\omega _{m}+\varepsilon _{m},%
\mathbf{k}+\mathbf{q}).  \label{ch_11}
\end{equation}
The computation is tedious but straightforward. The frequency-dependent part
of the  polarization operator for $q\ll k_{F}$ is given by
\begin{equation}
\Pi \left( \varepsilon _{m},q\right) =\frac{i\varepsilon _{m}}{2v_{F}q}\ln
\frac{i\varepsilon _{m}+v_{F}q}{i\varepsilon _{m}-v_{F}q}.  \label{pi3dm}
\end{equation}
Expanding $\Pi \left( \varepsilon _{m},q\right)$ in $\varepsilon _{m}/v_F q$,
we obtain
\begin{equation}
\Pi \left( \varepsilon _{m},q\right) =\frac{\pi \left| \varepsilon
_{m}\right| }{2v_{F}q}-\left( \frac{\varepsilon _{m}}{v_{F}q}\right) ^{2}+%
\mathcal{O}\left( \left| \varepsilon _{m}\right| ^{3}\right) .
\label{ps3dm_1}
\end{equation}
Substituting Eq.(\ref{ps3dm_1}) into Eq.(\ref{ch_11}), integrating over $q,$
and summing over $\varepsilon _{m}$ (which, to logarithmic
accuracy, is equivalent to just integrating over $\varepsilon _{m}$), we find that $%
\Sigma (\omega _{m},k)$ is  given by Eq.[\ref{se12}] with

\begin{equation}
\Sigma ^{(1)}(i\omega _{m}-\epsilon _{k})= \left( \frac{mk_{F}}{\pi ^{2}}%
U\right) ^{2}~\frac{(i\omega -\epsilon _{k})^{3}}{384E_{F}^{2}}~\ln \frac{%
E_{F}^{2}}{(i\omega -\epsilon _{k})^{2}}.  \label{10}
\end{equation}
This contribution originates from  the
first, leading term in the expansion of $\Pi (\varepsilon _{m},q)$ in $%
\varepsilon _{m}$  in (\ref{ps3dm_1}), and it comes from
internal bosonic frequencies that exceed the external one, i.e., from $%
|\varepsilon _{m}|>|\omega _{m}|$.
The second term depends on $\omega _{m}$ and $\epsilon _{k}$ separately
and has contributions from both the first and second
terms in Eq. [\ref{ps3dm_1}]. To  logarithmic accuracy,
\begin{eqnarray}
&&\Sigma ^{(2)}(\omega _{m},k)=- i~\left( \frac{mUk_{F}}{\pi ^{2}}\right) ^{2}%
\frac{\omega _{m}^{3}}{16E_{F}^{2}}~\ln \frac{E_{F}}{i\omega +\epsilon _{k}}
\notag \\
&&+\mathcal{O}\left[ \left( i\omega _{m}-\epsilon _{k}\right) \ln \frac{E_{F}%
}{i\omega _{m}+\epsilon _{k}}\right]  \label{11}
\end{eqnarray}
 (see comment \cite{comm_4}). Note that $\Sigma ^{(2)}(\omega
_{m},k)$~comes from the range of bosonic frequencies between $-|\omega _{m}|$
and $|\omega _{m}|$, i.e., relevant internal frequencies which are smaller
than the external one.

Comparing Eqs.(\ref{10}) and (\ref{11}), we see that for a generic ratio of $%
\epsilon _{k}/\omega _{m}$ and, in particular, for a particle on the Fermi
surface, where $\epsilon _{k}=0$, both contributions
behave as $\Sigma (\omega _{m},k_{F})\propto \omega _{m}^{3}\ln |\omega
_{m}| $; this result is in agreement with~Ref.\cite{Amit68}. However, only $%
\Sigma ^{(2)}(\omega _{m},k)$ actually gives rise to the $T^{4}\ln T$ term
in $\Xi , $ and hence to the $T^{3}\ln T$ terms in the entropy and the
specific heat. Indeed, the contribution of $\Sigma ^{(1)}(\omega _{m},k)$ to
$\Xi $ is
\begin{eqnarray}
&&T\sum_{\omega _{m}}\int d\epsilon _{k}~\Sigma ^{(1)}(\omega
_{m},k)G_{0}(\omega_m ,k)  \notag \\
&\propto &T\sum_{\omega _{m}}\int_{-E_{F}}^{E_{F}}d\epsilon _{k}(i\omega
_{m}-\epsilon _{k})^{2}\ln \frac{E_{F}^{2}}{(i\omega _{m}-\epsilon _{k})^{2}}
 \notag \\
&=&2T\sum_{\omega _{m}}\text{Re}\int_{0}^{E_{F}-i\omega _{m}}d\zeta \zeta
^{2}\ln \frac{E_{F}^{2}}{\zeta ^{2}},\text{ }\zeta \equiv \epsilon
_{k}-i\omega _{m}. \notag \\
\label{ch_12}
\end{eqnarray}
The resulting integral is determined by the upper cutoff of the integration,
 and is an  analytic function of $\omega _{m}$, hence gives
only an analytic contribution to the thermodynamic potential. On the other
hand, the convolution of $\Sigma ^{(2)}(\omega _{m},k)$ with the Green's
function yields
\begin{eqnarray}
&&T\sum_{\omega _{m}}\int d\epsilon _{k}~\Sigma ^{(2)}(\omega
_{m},k)G_{0}(\omega ,k)  \notag \\
&\propto &T\sum_{\omega _{m}}\omega _{m}^{3}\int \frac{d\epsilon _{k}}{%
\epsilon _{k}-i\omega _{m}}~\ln \frac{E_{F}}{\epsilon _{k}+i\omega _{m}}.
\label{ch_14}
\end{eqnarray}
To logarithmic accuracy, the integral over $\epsilon _{k}$ gives
\begin{eqnarray}
&&\int \frac{d\epsilon _{k}}{\epsilon _{k}-i\omega _{m}}~\ln \frac{E_{F}}{%
\epsilon _{k}+i\omega _{m}}=\int \frac{d\epsilon _{k}\epsilon _{k}}{\epsilon
_{k}^{2}+\omega _{m}^{2}}\ln \frac{E_{F}}{\epsilon _{k}+i\omega _{m}}  \notag
\\
&&+i\omega _{m}\int \frac{d\epsilon _{k}}{\epsilon _{k}^{2}+\omega _{m}^{2}}%
\ln \frac{E_{F}}{\epsilon _{k}+i\omega _{m}}  \notag \\
&=&i\pi \text{{sgn}}\omega _{m}\left[ \int_{|\omega _{m}|}^{E_{F}}\frac{%
d\epsilon _{k}}{\epsilon _{k}}+\ln \frac{E_{F}}{|\omega _{m}|}\right] =2i\pi
{\text{sgn}}\omega _{m}~~\ln \frac{E_{F}}{|\omega _{m}|} \notag \\
  \label{ch_15}
\end{eqnarray}
Substituting this into Eq. (\ref{ch_10}), evaluating the frequency sum using
\begin{equation}
T\sum_{\omega _{m}}\omega _{m}^{3}\ln \frac{E_{F}}{|\omega _{m}|}=\frac{2\pi
^{3}}{15}~T^{4}\ln \frac{E_{F}}{T},  \label{ch_16}
\end{equation}
and differentiating over $T$, we obtain for the entropy per particle
\begin{equation}
S(T)=-\frac{\pi ^{4}}{5}\left( \frac{mUk_{F}}{\pi ^{2}}\right) ^{2}\left(
\frac{T}{v_{F}^{\ast }k_{F}}\right) ^{3}\ln \frac{v_{F}^{\ast }k_{F}}{T}
\label{ch_17}
\end{equation}
This coincides with Eq. (\ref{S3dfinal_1}) expanded to second order, and
evaluated at $U(0) =U$, $U(2k_F) =0$ [in this approximation, $K_{c}=mk_{F}U/\pi ^{2}$%
, and $K_{s}=0$]. We also verified that the contributions
to $\Sigma ^{(2)}$ from $\varepsilon _{m}/q$ and $\varepsilon _{m}^{2}/q^{2}$
terms in $\Pi $ yield equal contributions to $\Xi $. This is consistent with
our earlier observation that non-analytic term in $\Xi $ comes from the
cross-product of the $O(\varepsilon _{m})$ and $O(\varepsilon _{m}^{2})$
terms in $\Pi $.

We see therefore that non-analytic term in $S(T)$ can indeed be obtained by
evaluating the self-energy first and then convoluting it with the fermionic
Green's function. However,  only a part of the self-energy
actually contributes to the nonanalyticity in $S(T)$. The term in the
self-energy that depends on $i\omega _{m}-\epsilon _{k}$ gives a
renormalization of the quasiparticle weight and does not lead to a
nonanalyticity in $S\left( T\right) .$

It is  instructive to comment here on the role of  forward scattering.
In 2D, forward scattering is special in the sense that its
contribution to the self-energy is comparable to that from backscattering~
\cite{Chubukov04}; both are of order $\varepsilon _{m}^{2}\ln |\varepsilon
_{m}|$. In fact, the 2D analogue of $\Sigma ^{(1)}(\varepsilon _{m},k)$
comes entirely from forward scattering.
In 3D, forward scattering is much less effective because of phase space
restrictions. One can show that in this case $\Sigma ^{(1)}(\varepsilon
_{m},k)$, evaluated to all orders in the perturbation theory in the
dimensionless coupling $u$,
is not restricted to forward scattering but rather comes from a
wide range of scattering angles. As we will discuss in the
next Section, there still exists
$\omega^3 \log \omega$ contribution to $\Sigma (\omega, k_F)$ specific to
 forward scattering in 3D, but this contribution is  exponentially
 small for a weak interaction.

\section{Comparison to earlier studies}

As we said in the Introduction, the analysis of the $T^{3}\ln T$ term in the
specific heat of a 3D Fermi Liquid has a long history. In this section we
compare our findings with the existing literature on the subject. We begin
with reviewing the $T^{3}\ln T$ nonanalyticities for the cases of
electron-phonon and fermion-paramagnon interactions, and then proceed with
the generic Fermi-liquid case.

\subsection{Electron-phonon interaction}

The electron-phonon coupling leads to nonanalyticities in the entropy;
in fact, the nonanalytic correction to the Fermi-liquid fixed point
was discovered first by Eliashberg \cite{Eliashberg60} in a study
of acoustic phonons coupled to fermions via a deformation potential.
Integrating out the phonons, one obtains 
a new contribution to the electron-electron interaction. For optical
phonons, or for acoustic phonons with a piezoelectric coupling,
this new interaction is frequency dependent on the scale of the Debye
frequency, but at low frequencies it just generates
an additional contribution to the electron vertex $\Gamma^k$,
so the results we have presented in previous sections carry over directly.

The case of an acoustic phonon-deformation potential, studied by Eliashberg,
presents a new feature: the effective interaction itself contains
a nontrivial $\varepsilon^2/q^2$ term  which leads to a new non-analytic 
contribution to entropy 
in $D=3$ (but not in $D=2$). We discuss this case in more detail.
Perturbation theory in the electron-phonon interaction is controlled by a small Migdal 
 parameter -- the ratio of the sound velocity $c$ to the
Fermi velocity $v_F$. To leading order in $c^2/v_F^2$,
 the thermodynamic potential can
be expressed as \cite{Eliashberg60,bardeen}
\begin{equation}
\Xi _{e-ph}=\frac{1}{2}\int\frac{d\varepsilon}{\pi}n_B(\varepsilon/T)
\int d^{3}q \text{Im} \left[\ln D^{-1}\left(\varepsilon ,q\right)\right],  \label{9a}
\end{equation}
where $D^{-1}$ is the inverse phonon propagator given by
\begin{equation}
D^{-1}\left( \varepsilon,q\right) =D_{0}^{-1}\left( \varepsilon,q\right) +
2g_{\text{ph}}^{2} c^2
q^{2}\Pi (\varepsilon,q).  \label{9}
\end{equation}
with bare propagator
\begin{equation}
D_{0}^{-1}\left( \varepsilon,q\right) =-\varepsilon^{2}+c^{2}
q^{2}.
\label{9b}
\end{equation}
Here $g_{\text{ph}}$ is
the dimensionless
effective electron-phonon coupling.
In a typical metal, $g_{\text{ph}}\sim 1$ \cite{LL}.
The expression for the thermodynamic potential $\Xi_{e-ph}$ is similar to
 the RPA expression for electron-electron interaction, c.f. Eq. (\ref{SN}),
 but there are two importand differences. Firs, the bare propagator 
 $D_0$  describes gapless excitations. Second, the second term of Eq.(\ref{9})
 contains an extra factor $q^{2}$. This factor 
 guarantees that the  $q=0$ phonon mode does not  affect the electrons
(``Adler principle''). These two differences combine to preserve the
 $T^2 \log T$ form of the nonanalytic correction to $\gamma$ in $3D$.
One can rearrange the $\text{Im}\ln\dots $
in Eq.(\ref{9a}) by factoring out
purely real quantities to get
\begin{equation}
\text{Im} \ln D^{-1}=\text{Im} \ln\left[1-\frac{\varepsilon^2}{(c^*)^2q^2} +
g_{*}^2 \Pi (\varepsilon,q)\right]
\label{Imln}
\end{equation}
where now  $(c^*)^2=c^2 + 2 g_{ph}^2\text{Re} \Pi(q,0)$
is the sound velocity renormalized
by the analytic terms in $\Pi$, and $g_*$ similarly is 
 the renormalized coupling, 
$g^2_* = 2 g^2_{\text{ph}}(c/c^*)^2$.
In analogy to the previous section, we now expand in powers of
$\text{Im} \Pi$. As before, the  term with one real particle-hole pair
 (i.e., of first order in $\text{Im} \Pi$) 
combines with the leading $\varepsilon^2/q^2$ term in the interaction vertex
to give a $T^3\ln T$ nonanalyticity. In addition, however, another
 $T^3 \log T$ emerges from the combination of $\text{Im} \Pi$ and 
$\sim\varepsilon^2/c^*{}^2q^2$ term from  the bare phonon propagator.
Such term does not exist for electron-electron interaction.
A simple analysis shows that 
this new contribution is larger by the inverse
Migdal parameter
$(v_F/c)^2$ than the contribution from the interaction vertex and hence is the dominant phonon $T^3 \log T$ contribution to the specific heat~\cite{comm_11}

Note that the sign of the phonon $T^3\ln T$ term is opposite to that
of the electron one, i.e., $d\gamma(T)/dT$ is positive for electron-phonon
contribution and negative for the electron-electron contribution

In $D=2$ (when both electrons and phonons are two-dimensional)
the $q$ dependence of the phonon propagator remains the same, but
the integration measure changes, and  $g_{\text{ph}}^{2}$ term
does not give rise to a nonanalyticity in $\Xi $. Only the  ``conventional'' 
term of order $g_{\text{ph}}^{4}$ gives rise to non-analyticity. From this
perspective, the electron-phonon problem in 2D does not
differ qualitatively from the electron-electron interaction.
In the case of mixed dimensionality, e.g., planes of electrons
embedded into a 3D elastic continuum,
more complex situations are possible.

One may also extract the ``phonon'' $T^3\ln T$ term in the entropy from the 
nonanalytic form of the electron self-energy
for the electron-phonon interaction $\Sigma_{\text{ph}}(\omega_m)\sim \omega_m^3\ln |\omega_m|$ \cite{LL},
in analogy with  Eq.(\ref{ch_10}). We re-iterate 
that  a direct relation
 between the nonanalyticities in the self-energy and entropy exists only for the case when the self-energy depends only 
on frequency.
This point will be important for the discussion of the zero-sound mode
contribution to the entropy
in Sec.\ref{sec:ZS}.

Finally, we note that the large phonon-induced nonanalyticity in the entropy
is not paralleled by a similar term in the spin susceptibility. This comes 
 about because  phonons
contribute directly  to the charge vertex, and only inderectly 
 to the spin vertex, at the subleading order in the Migdal parameter.
The basic reasoning for this was given long time ago  by Prange and Kadanoff
\cite{kadanoff}, who showed that in the Migdal approximation, 
 phonon contribution to the thermodynamic potential is
 confined to the Fermi surface,
and  does not change appreciably when the Fermi surface
position changes. As a result, 
an applied magnetic field, which shifts  spin up and spin down 
 Fermi surfaces,
does not change the phonon contribution to the thermodynamic potential,
 to leading order in the Migdal approximation. 
 Mathematically, 
this can be seen as follows: (i) The static spin suscepbility
$\chi_s (q)$  is given by the particle-hole bubble. 
 (ii) For free fermions, $\chi_s (q)$ comes from the states away from Fermi
 surface, if one integrates first over
 fermionic dispersion $\epsilon_k$,  while the contribution from 
the states near the Fermi surface
 vanishes because the poles of   
the  integrand 
as a function of dispersion $\epsilon_k$ reside in the same half-plane.
(iii) The renormalization of the spin susceptibility due to electron-phonon 
interaction  occurs via  self-energy 
insertions into the spin-polarization bubble. 
As the self-energy does not depend
on the electron momentum,  these  insertions
still leave the two poles of the integral over $\epsilon_k$ 
in the same half-plane, hence the integral over low-energies still
 vanishes.

The analysis beyond leading order in
 the Migdal parameter requires further analysis, not given
 here. However, for comparable electron-phonon and
electron-electron couplings, the phonon renormalization of $\chi_s$ is
 of the same order as the purely electron
one.

\subsection{Paramagnon model}

In the paramagnon model \cite{Doniach66}, the fermion-fermion interaction is approximated
by an interaction between fermions and  overdamped long-wavelength spin fluctuations
(paramagnons) The model has been extensively used to describe itinerant
electrons near a ferromagnetic instability. While the analysis is almost identical
to the analyses we have already presented, the  possibility tuning the model
through a ferromagnetic quantum phase transition
raises a new issue which requires discussion.

The RPA thermodynamic potential per unit volume in the paramagnon theory
has the same form as Eq.(\ref{SN}) but with slightly different
normalization factors; we  write it explicitly
here for convenience:

\begin{equation}
\Xi =\Xi _{0}+\frac{3}{2}\int\frac{d\varepsilon}{\pi}n_B(\varepsilon/T)
\int \frac{d^{D}q}{(2\pi )^{D}}%
~\ln [D_{S}^{-1}(\varepsilon,q)].  \label{oct_15_4}
\end{equation}
Here $\Xi _{0}$ is the thermodynamic potential of the
noninteracting fermions and
$D_{S}^{-1}(\varepsilon,q)=1+2g{\mathcal P}_{ph} (\varepsilon ,q)$ is the
inverse propagator of spin fluctuations, $g$ is the spin-fermion
coupling, and
${\mathcal P}_{ph}$ is the full polarization bubble
for noninteracting fermions. This model is therefore equivalent to the
previously studied RPA case, but with only the $L=0$ spin channel
interaction retained, so that the determinant [cf. Eq. (\ref{ev3d})]
may be written down
immediately. However, one extra piece of physics becomes important.
This theory can describe a ferromagnetic instability,
and indeed an  original
motivation for the model was the study of effects of
long wavelength spin fluctuations, which may be expected
to be important in nearly ferromagnetic materials. Near a ferromagnetic
instability, the dependence of the  {\em analytic} part of
$\Pi$ on $q$ becomes important.
Decomposing $\Pi$ into the analytic ($ const + q^2$) and nonanalytic
 ($\Pi_{\text{NA}}$) parts, we have
\begin{equation}
D_{S}^{-1}(\varepsilon,q)=(1+B)+\xi_0^2q^2+2g \Pi_{NA} (\varepsilon ,q)
\label{DRPA}
\end{equation}
with $B= 2g\Pi(0,0)\rightarrow -1$ as  the ferromagnetic transition
is approached.
We now proceed as before expanding in $\text{Im} \Pi_{NA}$ (representing real
particle-hole pairs) and $\text{Re}\Pi_{NA}$ (representing the nonanalytic
part of the dynamical interaction. The expansion contains inverse powers
of $1+B+\xi_0^2q^2$, so that the power counting of the momentum integrals
changes for $q>\xi=\xi_0/\sqrt{1+B}$ \cite{comm_c}. The nonanalyticities
we have discussed in this paper arise only from the Fermi-liquid regime
$T/v_F<q<\xi^{-1}$. As criticality is approached, $\xi$ diverges
and the
temperature window in which the $T^3lnT$ term 
may be observed becomes vanishingly small.
It is thus somewhat misleading to state that the $T^3\ln T$ nonanalyticities
become the critical nonanalyticities;
rather, they  should be regarded as a property of the Fermi-liquid
regime only, which is the ``quantum disordered regime" of the
ferromagnetic quantum phase transition.

With this proviso,
the previously given analysis can be applied directly.
We note here
that the original work \cite{Doniach66,larkin}
focussed on the three-pair contribution, obtained by treating
$\text{Re} \Pi_{NA}$ as a constant. In $3D$ one finds
\begin{equation}
\delta S^{(3)}\left( T\right) =\frac{\pi ^{6}}{20}
\left( \frac{B}{\Lambda}\right)^3~\left(\frac{T}{v_{F}^{\ast }k_{F}}\right) ^{3}\ln \frac{\Lambda v_{F}^{\ast
}k_{F}}{T},  \label{1_1}
\end{equation}
with $B=-2 g{\bar \Pi}(0,0)$ and $\Lambda = (\xi0/\xi)^2 = 1 +B$.
This expression coincides with Eq. (\ref{S3dfinal_1}) , if one considers
only spin contribution, neglects the backscattering term in $K_{s}$ and all $%
{\bar{\Gamma}}$ except for ${\bar{\Gamma}}_{s,0}$, and identify $B$ with the
Landau parameter $\Gamma _{s,0}$, such that ${\bar{\Gamma}}_{s,0}=B/(1+B)$.
Note that the narrowness of the ``Fermi-liquid" momentum regime appears
here as a dependence of the upper cutoff of the logarithm. Incorporating
the cutoff into the analytical $T^3$ terms, while mathematically justified,
obscures this physics.

The single-pair contribution,
overlooked in the paramagnon literature~\cite{Doniach66,larkin},
arises (as previously discussed) by including the
nonanalytic momentum dependence of $Re \Pi$ and is
\begin{equation}
\delta S^{(1)}=-\frac{3\pi ^{4}}{5}\left( \frac{B}{\Lambda}\right) ^{2}\left( \frac{T}{%
v^{\ast }_Fk_{F}}\right) ^{3}\ln \frac{\Lambda v_{F}^{\ast }k_{F}}{T}.  \label{1_2}
\end{equation}
This expression also coincides with Eq. (\ref{S3dfinal_1}), if one again
considers only
the spin contribution, but now takes into account only only
backscattering term in $K_{s}$, neglect all ${\bar{\Gamma}}$ except for ${%
\bar{\Gamma}}_{s,0}$, and again identify $B$ with the Landau parameter $%
\Gamma _{s,0}$.
Comparing the two non-analytic contributions to $\delta S(T)$, we find
that for negative $B$, i.e., positive $g$,
they have the same sign and become equal at $B\approx -0.55$. Further away from the
ferromagnetic instability (at smaller $1+B$), the single-pair contribution
dominates, while closer to the instability, the three-pair contribution is
larger.

The same situation holds in 2D. Away from criticality, there is only one
nonanalytic contribution to $S(T)$, from $\Pi ^{2}(\varepsilon _{m},q)$,
i.e, from a single-pair excitation with a frequency-dependent interaction.
It yields
\begin{equation}
\delta S(T)\propto \left( \frac{B}{\Lambda}\right) ^{2}\left( \frac{T}{v^{\ast }_F k_{F}}%
\right) ^{2}.  \label{1_2_1}
\end{equation}
Eq [\ref{1_2_1}] only applies for $T<v_F/\xi$; again as criticality is approached
the window of validity closes.

\subsection{Fermi Liquid}

The nonanalytic term in the entropy of a 3D Fermi Liquid arising from a 
generic  fermion-fermion interaction was considered in a number of publications in
60s and 70s~\cite
{Anderson65,Balian65,Engelsberg66,Amit68,Riedel69,Pethick73a,Pethick75}.
Recently, the $T^{3}\ln T$ term in $S(T)$ was reproduced via
multidimensional bosonization~\cite{Houghton98,fradkin}.

\subsubsection{Zero-sound mode}
\label{sec:ZS}

The very early studies~\cite{Anderson65,Balian65,Engelsberg66} focused on
the nonanalytic contribution due to the interaction with zero sound. As this
theme returns in later work (see, e.g., Ref.\cite{Coffey93}), it is
worthwhile to summarize the status of the problem here. The idea that the
interaction with zero sound
may lead to a nonanalyticity in the specific heat
was put forward by Anderson \cite{Anderson65}. Following
this suggestion, Balian and Fredkin \cite{Balian65} considered a
phenomenological model which treated the interaction with zero sound in
analogy with the electron-phonon interaction. However, an
unphysical choice of the interaction vertex (it
remained nonzero at $q\rightarrow 0,$ and thus did not satisfy the Adler principle)
 led Balian and Fredkin to the conclusion that the
nonanalyticity in the self-energy was very strong ($\omega _{m}\ln
\left| \omega _{m}\right| $). Later, Engelsberg and Platzman \cite
{Engelsberg66} pointed out that the zero-sound vertex vanishes as $q$ for $%
q\rightarrow 0$, as it does for the deformation-potential-type interaction with
phonons [cf. Eqs.(\ref{9a},\ref{9})].
They obtained an $\omega _{m}^{3}\ln \left| \omega _{m}\right|$ non-analyticity
in the self-energy at the Fermi surface, with
prefactor proportional to the difference of the zero-sound and Fermi
velocities  $c_{\text{zs}}-v_F>0$.
In 3D, this difference is exponentially small for weak interactions $u$:
$c_{\text{zs}}-v_F\propto e^{-1/u}$.
In 2D, the analogous analysis yields an
$\omega^2_{m}\ln \left| \omega _{m}\right|$ term in the self-energy, with  a  prefactor  of order $u^2$.

In this and all subsequent
publications, which employed 
the zero-sound nonanalyticity in the self-energy,
it was \emph{assumed }that this
automatically implies a nonanalytic entropy.
However, the two of us and collaborators have shown recently \cite
{Chubukov04,Chubukov05} that this assumption is, generally speaking,
incorrect, and that the interaction with zero sound in 2D does \emph{%
not} lead to a nonanalyticity in the
entropy to any order in the
 interaction. The most general argument is that the
 thermodynamic potential can be obtained
directly in Matsubara frequencies, by
expanding the Luttinger-Ward functional in the bare (fermion-fermion)
interaction.  The zero-sound mode simply does not arise in this formalism.
Recently, Catelani and Aleiner
\cite{aleiner} argued away the forward-scattering (and thus the zero-sound) contribution to
thermodynamics on the basis of gauge invariance of the Fermi-liquid kinetic equation.

On a more technical level, we come back to the argument that a direct relation
between the nonanalyticities in the self-energy and entropy exists only if fermions
interact with slow boson modes, so that the self energy
is approximately $k$-independent ("local").
As the zero-sound velocity is necessarily larger than the
Fermi velocity, this relation does not have to hold for this case.
The breakdown of this relation 
becomes evident on examination of the real-frequency analogue of Eq.(\ref{ch_10})
\begin{widetext}
\begin{equation}
\delta S=\frac{2}{\pi }~\frac{\partial }{\partial T}\left[ \frac{1}{T}%
~\int \frac{d^{D}k}{(2\pi )^{D}}~\int_{-\infty
}^{\infty
}d\omega \omega \frac{\partial n_{F}}{\partial \omega }\left\{ \text{Re}%
\Sigma(\omega ,k)\text{Im}G_{0}(\omega
,k)+\text{Im}\Sigma(\omega ,k)\text{Re}G_{0}(\omega
,k)\right\} \right], \label{feb5_1}
\end{equation}
\end{widetext}
where $G_0(\omega,k)=(\omega-\epsilon_k+i0^+)$ and $n_F$ is the Fermi function.
In a local theory,
the $k$ integral of the $\text{Re}%
\Sigma\text{Im}$$G_{0}$ term in  Eq.(\ref{feb5_1}) 
vanishes, and only $\text{Re}\Sigma$
contributes to $\delta S$. In a non-local theory, 
both the $\text{Re}\Sigma$ and the $\text{Im}\Sigma$
terms contribute. An
explicit computation shows\cite{Chubukov04,Chubukov05} that 
the zero-sound contributions from the two terms in Eq.(\ref{feb5_1})
cancel each other in 2D.
Although we have not performed such a calculation in 3D,
we believe that the result will be the same as in 2D, i.e.,
the zero-sound contribution vanishes.
This also follows from 
the simple argument
that in the RPA (ring) approximation in 3D, the entropy for the interaction with zero-sound is given
by Eq.(\ref{SN}), in which the argument of the logarithm is
 replaced by the inverse effective interaction
$U_{\text{eff}}(\varepsilon,q)$. Near the zero-sound pole, $U_{\text{eff}}(\varepsilon,q)\propto
e^{-1/u}q^2/\left(\varepsilon^2-c_{\text{zs}}^2q^2\right)$ and the zero-sound contribution to the entropy
coincides with that of a free bosonic mode. Accordingly,
the entropy is analytic and  scales as $T^3$\cite{LL,comm_7}.
 
\subsubsection{Fermion-Fermion interaction}

Detailed calculations of the self-energy and
$S(T)$ in generic models of interacting fermions have been performed
by Amit, Kane and Wagner~\cite{Amit68} and Pethick and Carneiro~\cite
{Pethick73a}, and we compare our results with theirs.

Amit and co-workers computed the self-energy near the mass shell,
expressed the result in terms
of Landau parameters keeping only the partial amplitudes with $L=0,1$, and
then used the resulting expression for the self-energy to compute the
$T^{3}\ln T$ correction to the entropy obtaining a result somewhat
similar to the one presented here.
There are however three qualitative distinctions between their results and
ours. First, Amit et al. argued that the non-analytic $\omega _{m}^{3}\ln
\left| \omega _{m}\right| $ term in the self-energy comes only from the
Landau damping term ($\varepsilon /q$) in the polarization bubble.
As we have shown, there are actually two contributions, one from the
Landau damping
($\varepsilon _{m}/q$) term in $\Pi (_{m},q)$ (considered by Amit and co-workers)
and another from $\varepsilon _{m}^{2}/q^{2}$ (overlooked by Amit).
Second, Amit et al. argued that the non-analytic self-energy at the mass
shell partly comes from the interaction with zero sound.
We showed in the previous subsection that this interaction
does not contribute to $T^{3}\ln T$ term in $S(T)$.
Third, Amit et al. argued that $2k_{F}$ scattering does not contribute to
thermodynamic potential to second order in the interaction. We found that
the contributions from small momentum scattering and from $2k_{F}$
scattering are identical at this order (up to  prefactors $U(0)$ and $U(2k_F)$).
We also note that the final result of Amit et. al.  for $\delta S(T)$ [Eqs.
(IV.17) from \cite{Amit68}b and (VI.8) and (VII.18) from \cite{Amit68}a is not identical
to the result we presented. Their combination of $A_{0}$
and $A_{1}$ does not reduce to the square of the backscattering amplitude,
and their overall factor in $\delta S(T)$ and the relative prefactor for the terms
with $\pi ^{2}/12$ in $K_{s}$ in (\ref{S3dfinal_1}) are different from ours.

Our result for the entropy fully agrees with that of Carneiro and Pethick~
\cite{Pethick73a}, if we use the same approximation as they did, namely,
neglecting all Landau coefficients with $L>2$. Within this approximation, our
Eq. (\ref{S3dfinal_1}) coincides with Eqs. (22) and (A19) in Ref.\cite
{Pethick73a}. In particular, to second order in the scattering amplitude,
their prefactor for the entropy is $%
A_{0}^{2}+A_{1}^{2}+A_{2}^{2}+2A_{0}A_{2}-2A_{2}A_{1}-2A_{0}A_{1}$ (the last
term is missing in (A19) of Ref.\cite{Pethick73a}, but this is obviously a
misprint). This combination is nothing but the square of the backscattering
amplitude $A_{\text{BS}}^{2}=(A_{0}-A_{1}+A_{2})^{2}$, which is the first
term in our prefactor $K$ in (\ref{S3dfinal_1}). From this perspective, our
new result is the observation that sum of all bilinear products of partial
components of $A$ reduces to the square of the backscattering amplitude.

It is frequently stated in the literature that Pethick and Carneiro considered
forward scattering between Landau quasiparticles, apparently
in disagreement with our result that  forward scattering
does not give rise to a nonanalyticity in $\Xi $.
There is in fact no contradiction because the forward scattering
considered by Pethick and Carneiro involves small angle scattering
between particles slightly away from the Fermi surface, whereas we consider
scattering of particles at the fermi surface. 

To see this explicitly we note that
Pethick and Carneiro used the Fermi liquid relation between the
entropy and the thermal correction to the quasiparticle energy
\begin{equation}
\delta S(T)=\sum_{\mathbf{p}}\Delta \epsilon _{\mathbf{p}}(T)\frac{\partial
n_{\mathbf{p}}^{0}}{\partial T},  \label{c_17}
\end{equation}
where $n_{\mathbf{p}}^{0}$ is Fermi function (a sum over spins is implicit),
expressed $\Delta \epsilon _{\mathbf{p}}(T)$ in terms of the Landau function
\begin{equation}
\Delta \epsilon _{\mathbf{p}}(T)=\sum f_{\mathbf{p},\mathbf{p}+\mathbf{q}%
}~n_{\mathbf{p}+\mathbf{q}}^{0},  \label{c_18}
\end{equation}
and argued that when $f$ is expanded in powers of $\mathbf{p\cdot }\mathbf{q}
$, the expansion contains a quadratic term $(\mathbf{p\cdot }\mathbf{q})^{2}$%
. This quadratic term gives rise to a $T^{3}\ln T$ contribution to $\Delta
S(T)$.

It is important to realize that the Landau function
$f_{\mathbf{p},\mathbf{p}+\mathbf{q}}$
describes the
interaction between physical quasiparticles, for which $\omega=v^*_F(|p|-p_F)$.
Accordingly, $f_{\mathbf{p},\mathbf{p}+\mathbf{q}}$ coincides, up to an
overall factor, with the fully renormalized vertex $\Gamma \lbrack (\mathbf{p%
},\omega ),(p+q,\omega +\varepsilon );(\mathbf{p}^{\prime }\omega ^{\prime
}),(\mathbf{p}^{\prime }+\mathbf{q},\omega ^{\prime }+\varepsilon )]$ (where
the first two pairs of arguments corresponds to the initial states and the
last two to the final ones), taken at $\mathbf{p}=\mathbf{p}^{\prime
},\omega ^{\prime }=\omega $ and $\varepsilon =\epsilon _{\mathbf{p}+\mathbf{%
q}}-\epsilon _{\mathbf{p}}\approx v_{F}\mathbf{n}_{\mathbf{p}}\cdot \mathbf{q%
}$, where $\mathbf{n}_{\mathbf{p}}$ is a unit vector along $\mathbf{p}$. In
fact, the $(\mathbf{p\cdot }\mathbf{q})^{2}$ term in $f_{\mathbf{p},\mathbf{p%
}+\mathbf{q}}$ describes how the interaction between quasiparticles evolves
when they move away from the Fermi surface. A similar consideration has
recently been applied to the analysis of nonanalytic terms in the spin
susceptibility~\cite{galitski05}.

The full irreducible vertex
\begin{widetext}
\begin{equation}
\Gamma \lbrack (\mathbf{p},\omega ),(p+q,\omega
+v_{F}\mathbf{n}_{\mathbf{p}}\cdot \mathbf{q});(\mathbf{p}^{\prime }\omega
^{\prime }),(\mathbf{p}^{\prime }+\mathbf{q},\omega ^{\prime }+v_{F}\mathbf{n%
}_{\mathbf{p}}\cdot \mathbf{q})]|_{\varepsilon =v_{F}\mathbf{n}_{\mathbf{p}%
}\cdot \mathbf{q}}=\Gamma _{\mathbf{p,p}^{\prime }}(\theta _{0}),
\label{aug21_2}
\end{equation}
\end{widetext}
 where $\theta _{0}$ is the angle between $\mathbf{p}$ and $\mathbf{q}$, satisfies
an integral equation, graphically shown in Fig. 1 of \cite{Pethick73a}.
The quadratic
dependence on $\cos \theta_0 $, which eventually leads to $T^3\ln 
T$ term in $S(T)$, comes from a virtual particle-hole bubble,
composed from quasiparticles at finite energy, i.e., away from the
Fermi surface. All analytic corrections are absorbed into the bare
vertex $\Gamma _{\mathbf{p,p}^{\prime }}^{0}(\cos \theta _{0})$,
in which one can then safely set $\cos \theta _{0}=0$. Explicitly,
\begin{equation}
\Gamma _{\mathbf{p,p}^{\prime }}(\theta _{0})=\Gamma _{p,p^{\prime
}}^{0}+A\int d^{3}p_{1}\Gamma _{\mathbf{p,p}_{1}}^{0}G_{\mathbf{p}_{1}}G_{%
\mathbf{p}_{1}\mathbf{+q}}\Gamma _{\mathbf{p}_{1}\mathbf{,p}^{\prime
}}(q,\theta _{0}),  \label{ch_18}
\end{equation}
where $A$ is a constant prefactor. The bare vertex describes the interaction
between physical fermions at the Fermi surface and coincides, up to an
overall factor, with $\Gamma ^{k}(\mathbf{n}_{p}\cdot \mathbf{n}_{p^{\prime
}})$ from Eq. (\ref{leg}). However, $\Gamma _{\mathbf{p,p}}(\theta )\propto
f_{\mathbf{p},\mathbf{p}+\mathbf{q}}$ is only obtained by solving the
integral equation (\ref{ch_18}).

Pethick and Carneiro solved Eq. (\ref{ch_18}) keeping only $L=0..2$ partial
components of $\Gamma ^{k}$ and found contributions to $\cos ^{2}\theta _{0}
$ term in $\Gamma _{\mathbf{p,p}}(\theta _{0})$ of the second and third
orders in $\Gamma _{\mathbf{p,p}^{\prime }}^{0}$. The second-order
contribution and a part of the third-order one correspond physically to
processes with one excited particle-hole pair, while the rest of the
third-order contribution describes the process with three excited
particle-hole pairs.

One can easily demonstrate, starting from Eq.(\ref{ch_18}), that the
second-order contribution is indeed expressed in terms of the backscattering
amplitude, as we found in Eq. (\ref{S3dfinal_1}). Applying one iteration to
Eq. (\ref{ch_18}) and integrating over $\epsilon _{\mathbf{p}_{1}}$ and over
frequency, we obtain
\begin{equation}
\Gamma _{p,p}(\theta _{0})\propto \cos \theta _{0}{}\int \frac{d\Omega
_{\theta }}{4\pi }\frac{\Gamma _{p,p_{1}}^{0}\Gamma _{p_{1},p}^{0}}{\cos
\theta _{0}-\cos \theta }  \label{ch_19}
\end{equation}
where $\Omega _{\theta }$ is a solid angle and $\theta $ is the angle
between $\mathbf{p}_{1}$ and $\mathbf{q}$. Substituting an expansion of $%
\Gamma _{\mathbf{p,p}_{1}}^{0}=\Gamma ^{0}(\theta -\theta _{0})$
\begin{equation}
\Gamma _{\theta -\theta _{0}}^{0}=\sum_{n}{\bar{\Gamma}}_{n}\cos ^{n}(\theta
-\theta _{0}),
\end{equation}
into Eq.(\ref{ch_19}), expanding the integrand to order $\cos ^{2}\theta
_{0} $ and evaluating the prefactors in the same way as in Sec. IV, we
obtain after rather straightforward algebra that
\begin{equation}
\Gamma _{p,p}(\theta _{0})\propto \cos ^{2}\theta _{0}\sum_{n,m}{\bar{\Gamma}%
}_{n}{\bar{\Gamma}}_{m}(-1)^{n+m}=\cos ^{2}\theta _{0}~\Gamma _{\text{B}%
S}^{2}
\end{equation}
This fully agrees with Eq. (\ref{S3dfinal_1}).

The same analysis can also be applied to the 2D case. The only difference is
that in 2D the $T^{2}$ term in $S(T)$ comes from all angles $\theta _{0},$
rather than from a specific range of small $\theta _{0}$. Still, one can
obtain an integral equation for $\Gamma _{\mathbf{p,p}^{\prime }}(\theta
_{0})$, similar to Eq.(\ref{ch_18}) and solve it by the same method as in
Sec. V.

\section{Comparison to  Experiment}
 
 In this section we present a brief comparison
of the results to measurements of the
$T^{2}\ln T$ term in the specific heat coefficient $\gamma (T)=C(T)/T$
in $^{3}$He~ \cite{Greywall83} and in several heavy-fermion materials \cite{Stewart94,Devisser87}. 
The $^3He$ data were analyzed using a formula equivalent to
our Eq. \ref{S3dfinal_1} but retaining only $L=0,1$ harmonics 
in  charge and spin channels.
The values of $\Gamma_{c,0}$, $\Gamma_{c,1}$, and $\Gamma_{s,0}$ were 
taken from independent measurements
of the compressibility, effective mass, 
and spin susceptibility, respectively, and $\Gamma_{s,1}$ was chosen as a free parameter
 to fit the observed $T^2\ln T$ term in$\gamma(T)$.  
At zero pressure, $\Gamma_{c,0}=9.15$,$\Gamma_{c,1}=5.27$, 
$\Gamma_{s,0}=-0.700$, and $\Gamma_{s,1}=-0.55$ \cite{Greywall83}. 
Use of Eq.(\ref{leg}) then gives $\Gamma_{\text{BS},c}=-1.01$ and $\Gamma_{\text{BS},s}=-1.66$, which yields for the 
combination of backscattering
amplitudes entering Eq.(\ref{S3dfinal_1}) $\Gamma2_{\text{BS},c}+3\Gamma2_{\text{BS},s}=9.28$.
Estimating the rest of the terms in $K_c+3K_s$ 
(which are cubic in the 
amplitudes) in the same approximation, we obtain $\left(K_c+3K_s\right)_\text{cubic}=20.05$. 
Thus in $^3He$ at ambient pressure it appears that the
backscattering contribution is about a third of the total result. At 
higher pressures, the relative importance of the 
backscattering contribution diminishes to about a quarter of the total result.
This suggests that in $^3$He
the "paramagnon" ($\Gamma_{s,0}^3$) contribution is quite important,
although the lack of information about higher Landau parameters
means that this conclusion must be regarded as provisional.

Similar analysis of the data for UPt$_3$ was performed in Ref.~\cite{Devisser87}. 
In this case, however, only the $L=0$ harmonics in Eq.(\ref{S3dfinal_1})
 were retained. The value of $\Gamma_{s,0}$
extracted from the magnitude of the $T^2\ln T$ term in $\gamma(T)$ does not agree 
with the value obtained from the enhancement of the spin susceptibility. Given this controversy,
we refrain from estimating the relative importance of the backscattering contribution to $\gamma(T)$ 
for this system. 
\section{Summary}

We have revisited the old subject of nonanalytic contributions to the
entropy (or specific heat coefficient) in light of recent progress in our
understanding the structure of the leading corrections to Fermi-liquid
theory. We have clarified the connection to the Fermi Liquid fixed point
by showing that the
coefficients of the nonanalytic terms may be expressed in terms of partial
harmonics of the fully reducible static scattering amplitude $\Gamma ^{k}(%
\mathbf{n}_{\mathbf{p}},\mathbf{n}_{\mathbf{p}^{\prime }})$ with both
momenta taken on the Fermi surface. We have presented a formalism
which allows us to distinguish between  excited particle-hole pairs
and their (possibly nonanalytic) interaction.

We have found two classes of contributions to the specific heat
nonanalyticity.  The first one, pertinent
to both two- and three-dimensional Fermi liquids, arises from an excitation
of a single-particle pair above the ground state combined
with a nonanalyticity in the quasiparticle interaction vertex.
The second one, pertinent only to
Fermi liquids in dimension $D\ge 3$,
arises from the excitation of three
particle-hole pairs, which interact via the analytic fixed-point interaction. In $D<3$ the first sort of
nonanalyticity is determined solely by the backscattering
amplitude, but in $D \ge 3$ additional contributions may arise.

The fact that the non-analyticity in the entropy involves the spin
component of the scattering amplitude suggests that the same physics is responsible
for the nonanalyticities in the susceptibility
which have been discussed extensively
elsewhere \cite{belitz,chubukov03,Chubukov04,Chubukov05}. We
have presented simple calculations which confirm this suggestion. We 
thus conclude that in
$D=3$ the nonanalytic momentum and magnetic field 
dependences of  $\chi_s$, reported earlier, involve terms
up to third order in the exact quasiparticle scattering amplitude, and the
quadratic term is the square of the spin component of the backscattering
amplitude.  From this perspective, the nonanalyticities
in the spin susceptibility are the same nonanalyticities
which have been studied in the context of specific heat for many years.

We acknowledge helpful discussions with I. Aleiner, I. Gornyi, P. Kumar, A. Melikyan, 
A. Mirlin, and G. Stewart, 
  and support from
NSF-DMR 0240238 (A. V. Ch.), NSF-DMR-0308377 (D. L. M.) and NSF-DMR-0431350
(A. J. M). 

\appendix

\section{Appendix A: $2k_{F}$ contribution to nonanalyticity in the specific
heat.}

We argued in  Sec. II that one can can re-express the $2k_{F}$
contribution to the thermodynamic potential $\Xi$ as an effectively small $q$
contribution. For completeness, however, we show in this Appendix how to
evaluate the $2k_{F}$ contribution to $\Xi$ directly.
The technical difficulty of this approach
is that one needs to know the polarization bubble near $2k_{F}$ at finite
frequency \emph{and }at finite temperature.
 For simplicity,
 we limit our analysis to the second-order perturbation theory.

We first obtain the asymptotic form of $\Pi \left( \varepsilon
_{m},q,T\right) $ and then use it in the calculation of the thermodynamic
potential.
It is convenient to separate $\Pi \left( \varepsilon _{m},q,T\right) $  into a
$T=0$ part and the $T-$dependent part as
\begin{equation*}
\Pi \left( \varepsilon _{m},q,T\right) =\Pi _{0}+\Pi _{T},
\end{equation*}
where
\begin{widetext}
\begin{eqnarray*}
\Pi_0\equiv\Pi\left( \varepsilon _{m},q,0\right)  &=&\int \frac{d^{3}k}{\left( 2\pi
\right) ^{3}}\Theta \left( -\epsilon _{\mathbf{k}}\right) \left[ \frac{1}{%
i\varepsilon _{m}-\epsilon _{\mathbf{k+q}}+\epsilon _{\mathbf{k}}}-\frac{1}{%
i\varepsilon _{m}-\epsilon _{\mathbf{k}}+\epsilon _{\mathbf{k-q}}}\right]  \\
\Pi_T\equiv\Pi \left( \varepsilon _{m},q,T\right)- \Pi \left( \varepsilon
_{m},q,0\right) &=&\int \frac{d^{3}k}{%
\left( 2\pi \right) ^{3}}\left[ n_{F}\left( \epsilon _{\mathbf{k}}\right)
-\Theta \left( -\epsilon _{\mathbf{k}}\right) \right] \left[ \frac{1}{%
i\varepsilon _{m}-\epsilon _{\mathbf{k+q}}+\epsilon _{\mathbf{k}}}-\frac{1}{%
i\varepsilon _{m}-\epsilon _{\mathbf{k}}+\epsilon _{\mathbf{k-q}}}\right] ,
\end{eqnarray*}
\end{widetext}
and $n_{F}\left( \epsilon _{\mathbf{k}}\right) =1/\left( \exp \left(
\epsilon _{\mathbf{k}}/T\right) +1\right) $ is the Fermi function. Consider
the $T=0$ part first. Expanding fermionic dispersions  near $q\approx
2k_{F}$ yields
\begin{equation*}
\epsilon _{\mathbf{k}\pm \mathbf{q}}=-\epsilon _{k}+v_{F}\bar{q}%
+2E_{F}\alpha ^{2},
\end{equation*}
where $\bar{q}=q-2k_{F}\ll 2k_{F}$ and $\alpha $ $\ll 1$ is the angle
between $-\mathbf{q}$ and $\mathbf{k}$ (for $\epsilon _{\mathbf{k}+\mathbf{q}%
})$ and between $\mathbf{q}$ and $\mathbf{k}$ (for $\epsilon _{\mathbf{k}-%
\mathbf{q}}).$ We assume that the $k$ integral is confined to the Fermi surface,
and replace $\int d^{3}k/\left( 2\pi \right) ^{3}\dots \rightarrow
\left( \nu _{F}/2\right) \int d\epsilon _{k}\int d\alpha \alpha \dots ,$
where $\nu _{F}=mk_{F}/2\pi ^{2}$ is the density of states at the Fermi
level per one spin orientation. Integrating over $\epsilon _{k},$ we arrive
at
\begin{equation*}
\Pi _{0}=\frac{\nu _{F}}{16}\int_{0}^{1}dx\ln \left[ \left( \bar{Q}+x\right)
^{2}+\bar{\Omega}^{2}\right] ,
\end{equation*}
where
\begin{equation*}
\bar{\Omega}_{m}\equiv \varepsilon _{m}/E_{F},\bar{Q}\equiv v_{F}\bar{q}%
/E_{F},\text{ and }x\equiv 2\alpha ^{2}.
\end{equation*}
Performing an elementary integration, we obtain
\begin{eqnarray}
&&\Pi _{0}=\frac{\nu _{F}}{16}  \label{2kfT=0} \\
&&\times \left[ \pi |{\bar{\Omega}_{m}}|-2|{\bar{\Omega}}_{m}|\arctan \frac{%
\bar{Q}}{|{\bar{\Omega}}_{m}|}-{\bar{Q}}\ln ({\bar{\Omega}}_{m}^{2}+{\bar{Q}}%
^{2})\right] .  \notag
\end{eqnarray}
Notice that since we restricted the momentum integration to the vicinity of
the Fermi surface, our expression does not contain a regular part of $\Pi ,$
which comes from all energies. However, in what follows we will need only a
nonanalytic part of $\Pi .$

 At $\varepsilon _{m}=0$, Eq.(\ref{2kfT=0})
reproduces the static Kohn anomaly
\begin{equation}
\Pi _{0}|_{\varepsilon _{m=0}}=-\frac{\nu _{F}}{8}{\bar{Q}}\ln {|{\bar{Q}}|}.
\label{th_2}
\end{equation}
At finite $\varepsilon _{m}$ and $q=2k_{F}$, i.e., ${\bar{Q}}=0$,
\begin{equation}
\Pi _{0}|_{q=2k_{F}}=-\frac{\nu _{F}}{16}|\bar{\Omega}_{m}|.  \label{th_3}
\end{equation}
This expression implies that the spectral weight of particle-hole pairs with
the center-of-mass momentum near $2k_{F}$ is proportional to the excitation
frequency.

For the  $T-$ dependent part, we find  after integrating over $\alpha ,$
\begin{equation}
\Pi _{T}=-\frac{\nu _{F}}{8}\int_{0}^{\infty }d\xi n_{F}\left( \xi
E_{F}\right) \ln \frac{{\bar{\Omega}}_{m}^{2}+\left( 2\xi +{\bar{Q}}\right)
^{2}}{{\bar{\Omega}}_{m}^{2}+\left( 2\xi -{\bar{Q}}\right) ^{2}},
\label{delta_pi}
\end{equation}
where $\xi =\epsilon _{k}/E_{F}.$

The $2k_{F}$ contribution to the thermodynamic potential, $\Xi _{2k_{F}},$
is given by
\begin{eqnarray}
\Xi _{2k_{F}} &=&- 3~NU^{2}_{2k_F} T  \label{xi2kf} \\
&&\times \sum_{\varepsilon _{m}}\int_{-1}^{1}d\bar{Q}\left[ \Pi
_{0}^{2}+2\Pi _{0}\Pi _{T}+\Pi _{T}^{2}\right] .  \notag
\end{eqnarray}
where $N$ is the number density. The choice of the cutoff in the $\bar{Q}$
integral does not affect the result to logarithmic accuracy. The rest of the
calculation is simplified by making two observations. First, the first term $%
\left( \left| {\bar{\Omega}}_{m}\right| \right) $ in Eq. (\ref{2kfT=0}) is
independent of ${\bar{Q}}$, while the rest of the terms in this equation as
well as $\Pi _{T}$ are odd in ${\bar{Q}}$. The square of the first term in
Eq. (\ref{2kfT=0}) does not lead to a nonanalyticity in $\Xi ,$ whereas the
cross-products of this term with the rest of $\Pi $ is odd in ${\bar{Q}}$
and thus vanish upon integration. Therefore, the first term can then be
safely eliminated from Eq. (\ref{2kfT=0}) which can then be  written as
\begin{eqnarray}
&&\Pi _{0}=-\frac{\nu _{F}}{16}  \notag \\
&&\times \left[ 2|{\bar{\Omega}}_{m}|\arctan \frac{\bar{Q}}{|{\bar{\Omega}}%
_{m}|}+{\bar{Q}}\ln ({\bar{\Omega}}_{m}^{2}+{\bar{Q}}^{2})\right] .
\label{pi0}
\end{eqnarray}
Second, typical $\xi $ in Eq.(\ref{delta_pi}) are of order $\bar{T}\equiv
T/E_{F}$, while the $\ln T$ dependence of the thermodynamic potential comes
from the region ${\bar{Q}}^{2},{\bar{\Omega}}_{m}^{2}\gg \bar{T}^{2}$. The
integral in Eq.(\ref{delta_pi}) can then be expanded in powers of $\xi $,
and a simple analysis shows that one needs to keep only terms of order $\xi $
and of order $\xi ^{3}$. With this simplification, $\Pi _{T}$ becomes
\begin{equation}
\Pi _{T}=-\frac{\pi ^{2}\nu _{F}}{12}\left[ \bar{T}^{2}\frac{\bar{Q}}{\bar{Q}%
^{2}+{\bar{\Omega}}_{m}^{2}}-\frac{16}{5}\pi ^{2}\bar{T}^{4}\bar{Q}\frac{{%
\bar{\Omega}}_{m}^{2}-\bar{Q}^{2}/3}{(\bar{Q}^{2}+\bar{\Omega}_{m}^{2})^{3}}%
\right] .  \label{th_7}
\end{equation}

There are four distinct $T^{4}\ln T$ contributions to $\Xi _{2k_{F}}$. The
first one comes from $\Pi _{0}^{2}$ and and is obtained by forming a
cross-product of the two terms in Eq.(\ref{pi0}). In the limit of $\left|
\bar{Q}\right| \gg |{\bar{\Omega}}_{m}|,$
\begin{equation}
\Pi _{0} \approx -\frac{\nu _{F}}{16} \left[2 \bar{Q} \ln |\bar{Q}|  + \pi |{\bar \Omega}_m| \right],
\end{equation}
and this combination results in a
nonanalytic term in $\Xi _{2k_{F}}$
\begin{equation}
\Xi _{2k_{F}}\propto T\sum_{\bar{\Omega}_{m}}|{\bar{\Omega}}%
_{m}|^{3}\int_{\left| \bar{\Omega}_{m}\right| }\frac{d{\bar{Q}}}{\bar{Q}}%
\propto T^{4}\ln \frac{E_{F}}{T}.  \label{th_4}
\end{equation}
Contrary to the $q=0$ contribution, however, the explicit $T-$dependence of $%
\Pi $ [entering via the second and third term in Eq.(\ref{xi2kf})] give
three more $T^{4}\ln T$ contributions in the thermodynamic potential. The
second and third contributions come from cross-products of the $T^{2}$ and $%
T^{4}$ terms in Eq.(\ref{delta_pi}) with $\Pi _{0}$. The logarithmic terms
here come, respectively, from $\int d{\bar{Q}}/{\bar{Q}}\rightarrow \ln {%
\bar{\Omega}}_{m}\rightarrow \ln T$ and from $T\sum_{\Omega _{m}}1/|\Omega
_{m}|\rightarrow \ln T$. The fourth contribution comes from the square of
the $T^{2}$ term in Eq. (\ref{delta_pi}). Each of these four contributions
is of order $T^{4}\ln T$, such that
\begin{equation}
\Xi _{2k_{F}}=-\frac{3\pi ^{4}}{8}\left( \nu _{F}U (2k_F)\right) ^{2}NT\left( \frac{%
T}{v_{F}k_{F}}\right) ^{3}\ln \frac{E_{F}}{T}\sum_{i=1}^{4}\gamma _{i}~
\label{th_8}
\end{equation}
Using
\begin{eqnarray}
&&T\sum_{\Omega _{m}}|\Omega _{m}|^{3}\ln |\Omega _{m}|\rightarrow -\frac{%
2\pi ^{3}}{15}T^{4}\ln \frac{E_{F}}{T},  \notag \\
&&T^{3}\sum_{\Omega _{m}}|\Omega _{m}|\ln |\Omega _{m}|\rightarrow \frac{\pi
}{3}T^{4}\ln \frac{E_{F}}{T},  \notag \\
&&T^{4}\sum_{\Omega _{m}}\frac{1}{|\Omega _{m}|}\frac{1}{\pi }T^{4}\ln \frac{%
E_{F}}{T},  \label{th_9}
\end{eqnarray}
we find
\begin{equation}
\gamma _{1}=-\frac{8}{45},~~\gamma _{2}=-\frac{8}{9},~~\gamma _{3}=\frac{14}{%
45},~~~\gamma _{4}=\frac{2}{9}.  \label{th_10}
\end{equation}
Collecting all contributions, we finally obtain
\begin{equation}
\Xi _{2k_{F}}=\frac{\pi ^{4}}{5}\left( \nu _{F}U (2k_F)\right) ^{2}NT\left( \frac{T%
}{v_{F}k_{F}}\right) ^{3}\ln \frac{E_{F}}{T}.  \label{th_11}
\end{equation}
It is instructive to compare this result with small $q$ contribution to
$\Xi$. We  recall that $\Pi \left(
\varepsilon _{m},q\right) $ for $\left| \varepsilon _{m}\right| \ll q\ll
k_{F}$ is given by Eq.(\ref{ps3dm_1}). The logarithmic singularity in $\Xi$ is
obtained by taking  the cross-product of the $\left| \varepsilon _{m}\right|
/q$ and $\left( \left| \varepsilon _{m}\right| /q\right) ^{2}$ terms in Eq.(%
\ref{ps3dm_1})
\begin{equation*}
\Xi _{0}\propto T\sum_{\varepsilon _{m}}\left| \varepsilon _{m}\right|
^{3}\int dqq^{2}/q^{3}\propto T^{4}\ln \frac{E_{F}}{T}.
\end{equation*}
Calculating the prefactor, we obtain
\begin{equation*}
\Xi _{q=0}=\frac{\pi ^{4}}{5}\left( \nu _{F}U (0) \right) ^{2}NT\left( \frac{T}{%
v_{F}k_{F}}\right) ^{3}\ln \frac{E_{F}}{T}
\end{equation*}
This almost coincides with Eq. (\ref{th_11}), the only difference is
 between the  prefactors $U^2(0)$ and
$U^2(2k_F)$.


\begin{thebibliography}{99}
\bibitem{comm_4}
 The $\ln {E_{F}/}(i\omega +\epsilon _{k})$ term in Eq.(\ref
{11}) is the sum of the $\ln E_{F}/|\epsilon _{k}|$ contribution from the $%
\varepsilon_{m}^{2}/q^{2}$ term in $\Pi (\varepsilon_{m},q)$ and the
$\ln {|\epsilon
_{k}|/(i\omega_{m}-\epsilon _{k})}$ contribution from the $|\varepsilon
_{m}|/q$ term in $\Pi \left( \varepsilon_{m},q\right) $.
\bibitem{comm_11} Note that at not extremely small $T$, 
this $T^3 \log T$ contribution is 
 still   smaller  than the purely phonon $T^3$ 
contribution to the specific heat
by a factor of $g^2_{\text{ph}}c/v_F\ln\omega_D/T$, where $\omega_D$ is the Debye frequency.
\bibitem{comm_c}  Recent calculations demonstrate~\cite
{belitz,chubukov03,pepin} that the momentum dependence of $D_S(q,0)$ is by itself
non-analytic in $q$ and approximating it by $q^{2}$ is only justified at
some distance away from the criticality.
\bibitem{comm_7} Note that in 2D, a similar consideration leads to a
non-analytic $T^2$  contribution which is of the same order as the
 nonanalyticity from electron-electron interaction. In this situation,
 an  additional analysis  is required to eliminate the zero-sound contribution
~\protect\cite{Chubukov04,Chubukov05}.
\end{thebibliography}
\end{document}